\documentclass[12pt]{article}

\usepackage{epsf}

\def\np#1#2#3{{\it Nucl. Phys.} {\bf B#1} (#2) #3}

\def\jhep#1#2#3{{\it JHEP} {\bf #1} (#2) #3}

\def\Ref#1{(\ref{#1})}
\def\plabel#1{\label{#1}}

\newcommand{\be}{\begin{equation}}
\newcommand{\ee}{\end{equation}}
\newcommand{\beq}{\begin{eqnarray}}
\newcommand{\eeq}{\end{eqnarray}}
\newcommand{\bea}[2]{\be\label{#2}\begin{array}{#1}}
\newcommand{\eea}{\end{array}\ee}

\def\lfig#1#2#3#4{
 \begin{figure}
 \refstepcounter{figure}
 \label{#4}
 \addtocounter{figure}{-1}
 \epsfxsize=#3
 \centerline{\epsfbox{#2}}
 {\bf \caption{{\rm #1}}}
 \end{figure}
}

\def\twofig#1#2#3#4#5#6{
 \begin{figure}
 \refstepcounter{figure}
 \label{#6}
 \addtocounter{figure}{-1}
 \centerline{\epsfxsize=#3\epsfbox{#2}\hspace{1cm}\epsfxsize=#5\epsfbox{#4}}
 {\bf \caption{{\rm #1}}}
 \end{figure}
}

\def\Nb{{\rm \bf N}}
\def\Rb{{\rm \bf R}}
\def\Cb{{\rm \bf C}}
\def\Zb{{\rm \bf Z}}

\def\rangl{\right\rangle   }
\def\langl{\left\langle  }
\def\({\left(}
\def\){\right)}
\def\[{\left[}
\def\]{\right]}
\def\p{\partial}

\def\11{1\!\! 1}

\def\hf{{1\over 2}}

\def\eps{\varepsilon}

\def\vp{\varphi}

\def\o{\omega }

   \def\CA {{\cal A}}
   
   \def\CC {{\cal C}}

   \def\CF {{\cal F}}

   \def\CL {{\cal L}}
   \def\CM {{\cal M}}

   \def\CP {{\cal P}}
   
   \def\CR {{\cal R}}
   
   \def\CT {{\cal T}}

\def\xp{x_{_{+}}}
\def\xm{x_{_{-}}}
\def\xpm{x_{_{\pm}}}

\def\tp{t_1}
\def\tm{t_{-1}}
\def\tpm{t_{\pm 1}}

\def\Pe{\Psi^{_{E}}}
\def\pse{ \psi^{_{E}} }

\def\Pepm{\Pe_{\pm }}

\def\psepm{\pse_{_{\pm }}}

\def\CC{{\tilde C}}
\def\DD{{D}}
\def\tDD{{\tilde D}}

\def\mub{\mu_{_B}}
\def\mul{\mu_{_L}}
\def\yl{y_{_L}}
\def\laml{\lambda_{_L}}
\def\mubc{\mu_{_{B,c=1}}}
\def\muc{\mu_{_{c=1}}}
\def\tauk{\tau_{0}(k,1)}

\def\ttau{\tilde\tau}
\def\ss{s}

\def\ef{e^{-{1\over R}X}}
\def\eg{e^{-{1-R\over R^2}X}}

\def\oR{{1\over R}}
\def\ZFZZ{Z^{_{FZZ}}}
\def\ZZZ{Z^{_{ZZ}}}
\def\ZD{Z^{c=1}_{\rm Dir}}

\def\SMQM{\Sigma^{_{MQM}}}
\def\SFZZ{\CM}
\def\SCFT{\Sigma}

\begin{document}
%
%

\title{
D-branes and complex curves in $c=1$ string theory}

\author{Sergei Alexandrov\thanks{email: S.Alexandrov@phys.uu.nl}
}

\date{}

\maketitle

\vspace{-0.9cm}

\begin{center}
\it  Institute for Theoretical Physics \& Spinoza Institute, \\
Utrecht University, Postbus 80.195, 3508 TD Utrecht, The Netherlands
\end{center}

\vspace{0.4cm}

\begin{abstract}
We give a geometric interpretation for D-branes in the $c=1$
string theory. The geometric description is provided by complex curves
which arise in both CFT and matrix model formulations.
On the CFT side the complex curve appears from the partition function
on the disk with Neumann boundary conditions on the Liouville field
(FZZ brane).
In the matrix model formulation the curve is associated with the profile
of the Fermi sea of free fermions.
These two curves are not the same. The latter can be seen as a certain
reduction of the former. In particular, it describes only $(m,1)$ ZZ branes,
whereas the curve coming from the FZZ partition function encompasses all $(m,n)$ branes.
In fact, one can construct a set of reductions, one for each fixed $n$.
But only the first one has a physical interpretation in the corresponding matrix model.
Since in the linear dilaton background the singularities associated with the ZZ branes degenerate,
we study the $c=1$ matrix model perturbed by a tachyon potential where the degeneracy disappears.
From the curve of the perturbed model
we give a prediction how D-branes flow with the perturbation and
derive the two-point bulk correlation function on the disk
with the FZZ boundary conditions.
\end{abstract}

\section{Introduction}

Recently, a considerable progress has been achieved in understanding of non-perturbative physics
in non-critical string theories.
The progress has been made both from the CFT and from the matrix model side.
On the CFT side, the breakthrough happened due to \cite{FZZbrane,ZZ} where
the relevant boundary conditions in Liouville theory have been discovered and 
some structure constants also have been computed. 
These boundary conditions preserve the conformal invariance and can be associated 
to D-branes in non-critical strings.
Then the modular properties with 
the boundary conditions of \cite{FZZbrane} have been considered in \cite{TeschnerBnd}, 
and other structure constants have been found in  \cite{Hos,TeschPons,Pons}.
In the matrix model formulation, some new non-perturbative 
effects have been calculated and identified with the ones coming from the D-branes
\cite{McGreevyKB,MARTINEC,KlebanovKM,MTV,KAK,SAmn,deBoer,AmJa,Boyar}.
All comparisons between the two approaches showed perfect agreement and 
even led to a new proposal for the matrix description of string theories with world sheet
supersymmetry, type 0A and 0B theories \cite{TakToum,DKKMMS,KMS}.
 
Most of non-perturbative effects found in the matrix models were shown 
to correspond to the Dirichlet boundary conditions on the Liouville field which
describe the so called ZZ branes \cite{ZZ}.
There is a two-parameter family of consistent boundary conditions of this type 
which are referred as $(m,n)$ branes.
But only the basic $(1,1)$ brane played a role in the recent analysis \cite{KlebanovKM,KAK}.
Later, however, it was found that in the $c=1$ string theory the one-parameter
set of $(m,1)$ branes can be relevant \cite{SAmn}.    

On the other hand, it is known that the Dirichlet boundary states can be obtained 
as a combination of the Neumann boundary states \cite{Hos,MARTINEC,TeschnerRim,Pons}, {\it i.e.},
there is some non-trivial relation between the ZZ and FZZ branes constructed in \cite{FZZbrane}.
For $c<1$ string theories, 
a nice geometric interpretation for this relation as well as for the origin of these D-branes
was found in \cite{Seib} and then 
obtained an explicit matrix model realization in \cite{KK}.
It was shown that the partition function of the FZZ boundary state gives rise to a Riemann surface.
Its singularities are associated to the ZZ branes whose partition functions 
can be obtained as line integrals 
of a certain one-form around non-trivial cycles on the curve.
Moreover, the curve appears to be the same as the one describing the ground ring of bulk operators
and the resolvent of the corresponding matrix model.

Regarding these results, it is natural to try to generalize them to the case of $c=1$ string theory,
which was the subject of the main interest in the study of 
tachyon condensation in non-critical strings. The present paper is supposed to fulfill this task.
Namely, we construct complex curves $\SFZZ$ and $\SMQM$
coming from the FZZ partition function and from 
Matrix Quantum Mechanics (MQM), respectively. 
We show how they are related to each other and how they
incorporate non-perturbative physics. 

There are two main problems which appear when one generalizes the results of \cite{Seib}
to the $c=1$ case.
The first one is that the complex curve of \cite{Seib} degenerates in the limit $c\to 1$
in the sense that all singularities collapse to two points.  
Due to this reason and to find some new interesting information, 
we study the curve of the $c=1$ string theory perturbed by the Sine--Liouville potential,
which allows to distinguish between different ZZ branes.
On the MQM side the exact solution of the perturbed theory is known \cite{AKK},
whereas in the CFT formulation we can perform calculations only 
in the first order in the perturbation. To this order we find agreement
with the matrix model results and we suppose that it holds to be true
to all orders. With this assumption, the exact solution of the perturbed MQM allows us 
to make some predictions concerning disk correlation functions in the $c=1$ string theory.   
 
The second problem is that, unlike the $c<1$ case, the complex curve $\SFZZ$ following from 
the FZZ partition function is not quite the same as the curve $\SMQM$ appearing in MQM and which
was shown to describe the ground ring of the $c=1$ string theory \cite{Witgr}. 
The latter (in the non-perturbed case) is given by the standard hyperboloid, whereas
the former topologically coincides with its universal covering.
We will show that the relation between them 
is similar to the relation between the resolvent and the density
in the matrix model represented by a matrix chain. 
This relation appeared recently in the study of the boundary
ground ring of the $c=1$ string \cite{KostovGR}.

The fact that the two curves $\SFZZ$ and $\SMQM$ are not the same 
has an important consequence. Whereas the curve $\SFZZ$ is able to incorporate
all $(m,n)$ ZZ branes, the hyperboloid describes only a subset of them, the $(m,1)$ branes.
Thus, only this subset seems to be physically relevant in the $c=1$ theory. This fact was noted
previously within the MQM formulation in \cite{SAmn}.  

The paper is organized as follows. In the next section we construct the complex curve $\SMQM$ 
which describes the solution of MQM perturbed by a tachyon potential.
Its real section coincides with the profile of the perturbed Fermi sea
of the singlet sector of MQM, whereas the orthogonal section gives fermionic trajectories 
in Euclidean time. We show that the classical one-fermion action evaluated on  
closed Euclidean trajectories reproduces the non-perturbative contributions to the free energy
calculated in \cite{KAK,SAmn} and given by the disk partition functions of the $(m,1)$ ZZ branes. 
Each closed trajectory is associated with a singularity of $\SMQM$ where the curve has 
a self-intersection. In \cite{Seib} such singularities  were interpreted as pinched cycles.

Then in section \ref{CFTsection} we analyze complex curves arising in the CFT formulation.
First, we construct the complex curve $\SFZZ$ coming from the FZZ partition function by
taking the limit $c\to 1$ of the curve considered in \cite{Seib,KK}. All basic conclusions 
of \cite{Seib}, such as the D-branes are given by line integrals and the ZZ branes are associated 
to non-trivial closed cycles passing through the singularities of $\SFZZ$, remain to be true.
Second, we show what one needs to do with $\SFZZ$ to get a complex curve $\SCFT_1$ coinciding
with the matrix model curve $\SMQM$. The equivalence of the matrix model and the CFT constructions,
namely that $\SCFT_1=\SMQM$,
which is supposed to be true in the full perturbed theory, 
is proven to the first order in the Sine--Liouville coupling. 

After that we demonstrate that all singularities of $\SCFT_1$ correspond 
to the $(m,1)$ branes and derive a generalization of the relation between the FZZ and ZZ 
partition functions. In particular, we also predict how the positions of the $(m,1)$ branes
on the Riemann surface $\SFZZ$ change with the Sine--Liouville coupling.
Then we show that one can actually 
construct a Riemann surface $\SCFT_n$ for each set of $(m,n)$ branes 
with fixed $n$. But at the same time we argue that the perturbation seems to remove 
the singularities of $\SCFT_n$ and, therefore, only the $(m,1)$ branes remain in the theory.
Finally, we give some comments on an interesting symmetry of $\SCFT_1$ under rotations.
This symmetry has a clear origin in the discrete translational symmetry of the Sine--Liouville action,
but it can still have non-trivial implications for the non-perturbative physics. 

We conclude with discussion of the found results. In Appendix \ref{B} the reader can find
a new result on the two-point correlation function of the Sine--Liouville operator
on the disk with the FZZ boundary conditions derived from the MQM complex curve.

\section{Geometric description of instanton contributions in the perturbed MQM}

\subsection{MQM solution of Sine--Liouville theory}

We start from a review of the matrix model description of the $c=1$
string theory in non-trivial tachyonic backgrounds. (For a nice review of the $c=1$
string theory we refer to \cite{KLEBANOV,Moore}, a recent review on MQM is 
\cite{SAthese}.)
Mostly we will be interested in a particular case of such backgrounds which
is described by the so-called Sine-Liouville theory.
The CFT action in this case contains, besides the usual Liouville term,
one vertex operator of momentum $1/R$: 
\be
S_{\rm SL}={1\over 4\pi}\int d^2\sigma\left[(\partial X)^2 +(\partial\phi)^2
+2\hat \CR\phi+\mul \phi e^{2\phi}+\laml e^{(2-\oR)\phi}
\cos \(X /R \)\right].
\label{conepert}
\ee

As it is well known, the non-perturbed $c=1$ string theory is equivalent
to the singlet sector of MQM, which provides a powerful description
in terms of free fermions in the inverted oscillator potential.
A way to introduce tachyon perturbations in the fermionic picture
was suggested in \cite{AKK}.
It essentially uses the light cone representation of MQM based
on the light cone like coordinates in the phase space of fermions
\be
\xpm={x\pm p \over \sqrt{2}} .
\label{lccor}
\ee
It was shown that the tachyon perturbations correspond to
deformations of one-fermion wave functions in this representation
\beq
& \Pepm(\xpm)=e^{\mp i \vp_\pm(\xpm;E)}
\psepm(\xpm), & \label{asswave}
\\
& \vp_\pm  (\xpm;E)= V_\pm(\xpm) +\hf \phi(E) + v_\pm(\xpm;E), &
\label{fasewave}
\eeq
where
\be
\psepm(\xpm)= {1\over\sqrt{2\pi }}\xpm^{ \pm i E-\hf}
\label{wavef}
\ee
are non-perturbed eigenfunctions of the inverse oscillator Hamiltonian
\be
\hat H_0 = -\hf (\hat \xp\hat \xm +\hat \xm\hat \xp).
\label{oneph}
\ee
The perturbation is completely defined by the asymptotic part $V_\pm$ of
the perturbing phases. The rest contains the zero mode $\phi(E)$ and the part
$v_\pm$ vanishing at infinity. They are to be found from
some compatibility condition and expressed through
the parameters of the potentials $V_\pm$.

A special class of perturbations is given by the potentials $V_\pm$ of
the following form
\be
V_\pm(\xpm)
= \sum\limits_{k\ge 1} t_{\pm k} \xpm^{k/R} .
\label{Vbig}
\ee
It was proven that in this case the perturbed system is described by
a constrained Toda hierarchy \cite{AKK}. This fact allows to apply
the technique of integrable systems and to find an explicit solution.
In the simplest case, when only the first coupling constants, $\tpm$,
are non-vanishing, the perturbed MQM gives a description of the Sine-Liouville
theory \Ref{conepert}.

In the quasiclassical limit the system is described by the Fermi sea
made of the free fermions. All dynamical information can be extracted from
its exact form which is given by a curve in the phase space.
The position of the curve is defined by the compatibility of the following two
equations
\be
\xp\xm=\mu+W_{\pm}(\xpm;\mu)=
 {1\over R}\sum\limits_{k\ge 1}  k t_{\pm k} \,  \xpm^{ k/R}  +\mu  +
{1\over R}\sum\limits_{k\ge 1} v_{\pm k}(\mu)\,  \xpm^{-k/R},
\label{eqfs}
\ee
where $W_{\pm}=\p_{\pm}\vp_\pm$ and $\mu$ is the Fermi level.
The solution can be presented in the parametric form.
We restrict ourselves only to the case of the Sine-Liouville potential
with $\tp=\tm=\lambda$.
Then one obtains
\beq
& \xpm(\o) = e^{-{1\over 2R}\chi}  \o^{\pm 1} (1+ a \, \o ^{\mp {1\over R}} ),
\label{xpmo} & \\
& \mu e^{ {1\over R} \chi} -{1\over R^2}
\left(1-{1\over R}\right)\lambda^2 e^{{2R-1\over R^2} \chi} =1,
\qquad
a = {\lambda\over R} \, e^{{R-1/2\over R^2} \chi}. &
\label{param}
\eeq
The equation \Ref{eqfs} appears in the formalism of Toda hierarchy
as one of the constraints (string equation).
Another constraint is the requirement that
$\xpm(\o,\mu)$ describe a canonical transformation
from the canonical set of variables $(\log \o, \mu)$
\be
\{\xm,\xp\}=1 \Leftrightarrow \{\mu,\log\o\}=1.
\label{cantr}
\ee
Thus, the symplectic structure on the phase space is induced
by the commutation relations on the Toda lattice.

The quantity $\chi$ in \Ref{xpmo}
is related to the spherical part of the grand canonical free energy
by $\chi=\p^2_{\mu}\CF_0$. The string coupling $g_{s}$, which generates
the genus expansion of the free energy, can be associated
either with $\mu$ or $\lambda$. Let us choose the second possibility.
Then it is convenient to introduce the following scaling variables
\be
y=\mu\xi, \qquad \xi=\({1-R\over R^3}\,\lambda^2\)^{-{R\over 2R-1}} \sim g_s^{-1}.
\label{scpt}
\ee
Here we imply the restriction $1/2<R<1$. We need to impose it because the solution
is not analytic in $R$ and has two critical points corresponding to the self-dual
radius $R=1$ and the radius $R=1/2$ of the Kosterlits--Thouless phase transition.
We choose the interval containing T-dual of the black hole point $R=3/2$ \cite{FZZ,KKK}.
However, the most of our results are also true for $R>1$, possibly, with
modification $(1-R) \to |R-1|$.
In terms of the variables \Ref{scpt} the equation \Ref{param} for the free
energy can be rewritten as \cite{KKK}
\be
 y=\ef-\eg, \qquad   \chi= R\log\xi+X(y).
\label{wf}
\ee

Note that after a T-duality transformation, all results found in this way
coincide with the corresponding results for the system perturbed by windings \cite{KKK,AK}.
For the general potential \Ref{Vbig}, the T-duality transformation reads as follows
\be
R\to 1/R, \qquad \mu\to R\mu,\qquad
t_n\to R^{-nR/2}t_n.
\label{tdual}
\ee
The scaling parameters introduced in \Ref{scpt} transform under the T-duality as
\be
\label{scptr}
\xi\to \xi/R, \qquad y\to y.
\ee

\subsection{Leading instanton contributions in SL theory}
\label{leadinst}

The instanton corrections to the free energy of MQM with
the Sine--Liouville perturbation were investigated in \cite{KAK,SAmn}.
In that papers a T-dual model was considered where the theory was compactified
on a time circle of radius $R$ and, instead of the tachyon, a winding
perturbation was introduced. However, the results found in \cite{KAK,SAmn}
can be immediately translated to our case applying the T-duality transformation
\Ref{tdual} or \Ref{scptr}. As a result, we have the following picture.

In the $c=1$ matrix model there is an infinite set of
non-perturbative corrections labeled by $k\in \Nb$.
They appear in the formula for the non-perturbed free energy
and are given by $e^{-2\pi k \mu}$ \cite{BRKA,GIZI,PARISI,GRMI,Kazakov:1990ue}.
If the theory is compactified on a radius $R$, there appears
an additional set of corrections of the form $e^{-2\pi k R \mu}$ \cite{GK}.
These two sets have a natural interpretation from the CFT point of view
as contributions of string disk amplitudes with the Dirichlet and Neumann
boundary conditions on the matter field $X$, respectively. In both cases
a Dirichlet boundary condition on the Liouville field is imposed.
It was argued \cite{SAmn} that this should be the $(k,1)$ or $(1,k)$
boundary condition of Zamolodchikovs \cite{ZZ}.

After a tachyon perturbation, the second set disappears, whereas the corrections
from the first set become dependent on the parameters of the perturbation.
In the case of the Sine--Liouville perturbation,
to the leading order they have the following form
\be
\eps_k(\mu,y)\sim e^{-\mu f_k(y)}~.
\plabel{anz}
\ee
where the functions $f_k(y)$ are given by
\beq
&
f_k(y)=2\phi_k(y)+ {2(2R-1)\over y\sqrt{\oR-1}}
e^{-{1\over 2R^2} X(y)}\sin \(\oR \phi_k(y)\), &
\plabel{fphi}
\\
&
{ \sin \( \phi_k\)
}
=a(y)\sin \( {1-R\over R} \phi_k\) &
\plabel{solz}
\eeq
with the same parameter $a$ as in \Ref{param}.
Note that the functions $f_k$ have the same parametric form for all $k$ and
the functions $\phi_k(y)$ are directly related to the derivative of $f_k(y)$:
\be
\phi_k(y)=\hf \p_y(yf_k(y)).
\plabel{fph}
\ee
Different corrections \Ref{anz} are distinguished by different
solutions of the equation \Ref{solz}. The solutions are evident in the limit
where the perturbation is absent.
Therefore, they can be characterized by the initial conditions
at $\lambda=0$
\be
\lim\limits_{y\to\infty}\phi_k(y)=\pi k.
\plabel{ink}
\ee
Their explicit expansion in $\lambda$ reads as follows
\be
\phi_k(y) \approx
\pi k +{1\over R} \sin\(\pi k \over R\)\, \mu^{-{2R-1\over 2R}} \lambda
+{1-R\over 2R^3}\sin\(2\pi k\over R\)\, \mu^{-{2R-1\over R}}\lambda^2 +O(\lambda^3).
\label{corr}
\ee

Let us now reproduce these results from the analysis of the fermionic tunneling
which will provide a natural interpretation for
the parameters $\phi_k$.
In the spherical approximation a fermion propagating over
the Fermi sea is decoupled from it.
Therefore, to this order the perturbed system can be reformulated as
a fermion moving in the perturbed potential at the Fermi level.
The perturbed Hamiltonian is defined as a consistent solution
of the following equations \cite{AKK}
\be
H(\xp,\xm)=-\xp\xm+W_{\pm}\({\xpm,-H(\xp,\xm)}\)
\label{fHpert}
\ee
with the potentials $W_\pm(\xpm;\mu)$ defined in \Ref{eqfs}. 
The classical action can be written as
\be
\label{actpert}
S_{\rm ferm}=\int dt \(p\dot x -H(\xp,\xm)-\mu\).
\ee
The last term appears due to the passing to the grand canonical ensemble
of MQM and fixes the energy of the fermion to be equal to $-\mu$.

The classical trajectories of this system, which are solutions of the equation
\be
H(\xp,\xm)=-\mu,
\label{Hmu}
\ee
given in the parametric form in \Ref{xpmo}.
It turns out that the parameter $\o$ is directly related to
time $t$.\footnote{Note that this time does
not coincide with time of MQM in the initial
interpretation where we perturbed the wave functions (the state of the system)
instead of the Hamiltonian.
The dependence of that time is always described through
$\xpm(t)=e^{\pm t}\xpm(0)$.}
Indeed, the equations of motion read
\be
\label{eqmot}
\dot\xpm=\{\xpm,H \}=\p_{\log\o}\xpm,
\ee
where we used \Ref{Hmu} and \Ref{cantr}.
Therefore, one can identify
\be
\label{time}
t =\log\o.
\ee

When we quantize the theory, the classical trajectories \Ref{xpmo}
give the leading contribution to the path integral.
At the same time, the instanton contributions are related
to classical trajectories of the system analytically continued
to imaginary time $t\to -i\tau$. The trajectories are still described
by \Ref{xpmo}. However, the parameter $\o$ is not real anymore.
The relation \Ref{time} implies that it should be a pure phase $\o=e^{-i\tau}$.
This is consistent with the fact that in imaginary time $\xpm(\tau)$
should be complex conjugated to each other
since the momentum $p={\xp-\xm\over \sqrt{2}}$
becomes pure imaginary.

The crucial fact is that the functions $\phi_k(y)$ are exactly the moments of
Euclidean time $\tau$ when the momentum $p$ vanishes.
Indeed, taking $\o_{\pm k}=e^{\mp i\phi_k}$, one finds that
$\xp(\o_{\pm k})=\xm(\o_{\pm k})$.
Moreover, since $x(\o_{k})=x(\o_{-k})$,
we see that the Euclidean trajectory has self-intersections at the points
corresponding to these values of the parameter
\be
x_k=\sqrt{2} e^{-{1\over 2R}\chi}\(\cos\phi_k+ a\cos\({1-R\over R}\phi_k\)\)  ,
\qquad p_k=0.
\label{points}
\ee
Thus, varying $\tau$ in the interval $[-\phi_k,\phi_k]$, one obtains
a closed curve on the plane $(x,ip)$
with $2k$ intersections of the line $p=0$.
Note that the most right point of each curve is
$x_0=\sqrt{2} e^{-{1\over 2R}\chi}\(1+a\)$, whereas
the most left point, in general, does not belong to the axis
$p=0$.

With each closed curve one can associate an instanton contribution to
the free energy.
The leading contribution is given by the Euclidean
action evaluated on the classical trajectory,
$\CA_{\rm inst}\sim \exp\(-S^{\rm E}_{\rm ferm}\)$.
In our case one obtains
\be
\label{SEucl}
S^{\rm E}_{\rm ferm}(k{\rm -inst})=\int\limits_{-\phi_k}^{\phi_k}
d\tau \(-ip \p_{\tau} x +H(\xp,\xm)+\mu\)
=i\int\limits_{-\phi_k}^{\phi_k} \xm \p_{\tau}\xp d\tau=
2 \int\limits^{\mu} \phi_k d\mu,
\ee
where we took into account the equation \Ref{Hmu} valid on the classical trajectories
and used the canonical transformation \Ref{cantr}.
The boundary terms coming from the kinetic term disappear
since the trajectories are closed. Due to the relation \Ref{fph}
the last integral reproduces $f_k(y)$ from \Ref{fphi}.\footnote{The relation of 
the Euclidean action to the non-perturbative corrections was first
suggested by I. Kostov and proven in the first orders in $\lambda$ in \cite{KK}.}

Besides, we note that the T-dual instanton contributions can be obtained from 
non-closed trajectories going from $\tau=0$ to $\tau=2\pi k R$
\be
\label{SEuclT}
S^{\rm E}_{\rm ferm}(\tilde k{\rm -inst})=\int\limits^{2\pi k R}_{0}
d\tau \(-ip \p_{\tau} x +H(\xp,\xm)+\mu\)
=2\pi k R \mu
\ee
what follows from the explicit relation \Ref{eqfs} between $\xp$ and $\xm$.
This fact will get a geometric interpretation in the next subsection.

Finally, to complete the picture, we mention that
the same classical action evaluated on the real trajectory
$\xpm(\o),\ \o\in \Rb$ reproduces the perturbative part of the free energy \cite{AKKNMM}.

\subsection{Complex curve of the perturbed $c=1$ matrix model}
\label{compcurv}

The analysis of the previous section shows that it is natural to consider
a complex curve defined by the equation \Ref{Hmu} where now the light-cone
coordinates $\xpm$ are considered as complex variables. Let us call it $\SMQM$.
Due to \Ref{SEucl} and \Ref{SEuclT} the instanton contributions are given by integrals
along various cycles on this curve.
This result, together with the analysis of string theories with the central
charge $c<1$ \cite{Seib,KK}, indicates that the curve should have
an interpretation in terms of D-branes.
We will reveal such interpretation in the next section.
Here we will describe the main properties of the complex curve \Ref{Hmu}
as they appear in the matrix model.

The curve can be thought as a surface in $\Cb^2$ with the flat coordinates $(\xp,\xm)$.
Explicitly, it is described by the functions $\xpm(\o)$ given in \Ref{xpmo}
where $i\log\o=\tau\in \Cb$.
Thus, $\tau$ is a uniformization parameter of this Riemann surface.
However, if we are interested in the properties of its embedding into $\Cb^2$,
they are quite non-trivial.

Let us first consider the complex curve of the non-perturbed theory.
It represents the standard hyperboloid
\be
x^2-p^2=\mu.
\label{hyperb}
\ee
Since in the parameterization
\be
x=\sqrt{\mu}\cosh(i \tau), \qquad p=-\sqrt{\mu}\sinh(i \tau)
\label{paramnon}
\ee
we allow for the parameter $\tau$ to run over the whole complex plane,
the complex curve actually wraps the hyperboloid infinitely many times.

Due to this infinite wrapping the arising picture is quite degenerate.
As we will see, the perturbation removes the degeneracy.
To understand the picture after the perturbation,
we analyze the properties of the complex curve $\SMQM$ in the initial coordinates
$(x,p)$. We will distinguish two cases: when
the parameter $R$ is rational or not.

\subsubsection{Rational $R$}

\twofig{The case of a rational value of the radius:
$R=2/3$. a) Section of the complex curve $\SMQM$ by the plane \Ref{planez} giving
a trajectory in Euclidean time. b) Section of the curve by the plane \Ref{planex}
giving the profile of the Fermi sea and a finite number of points
on the $p=0$ axis.}{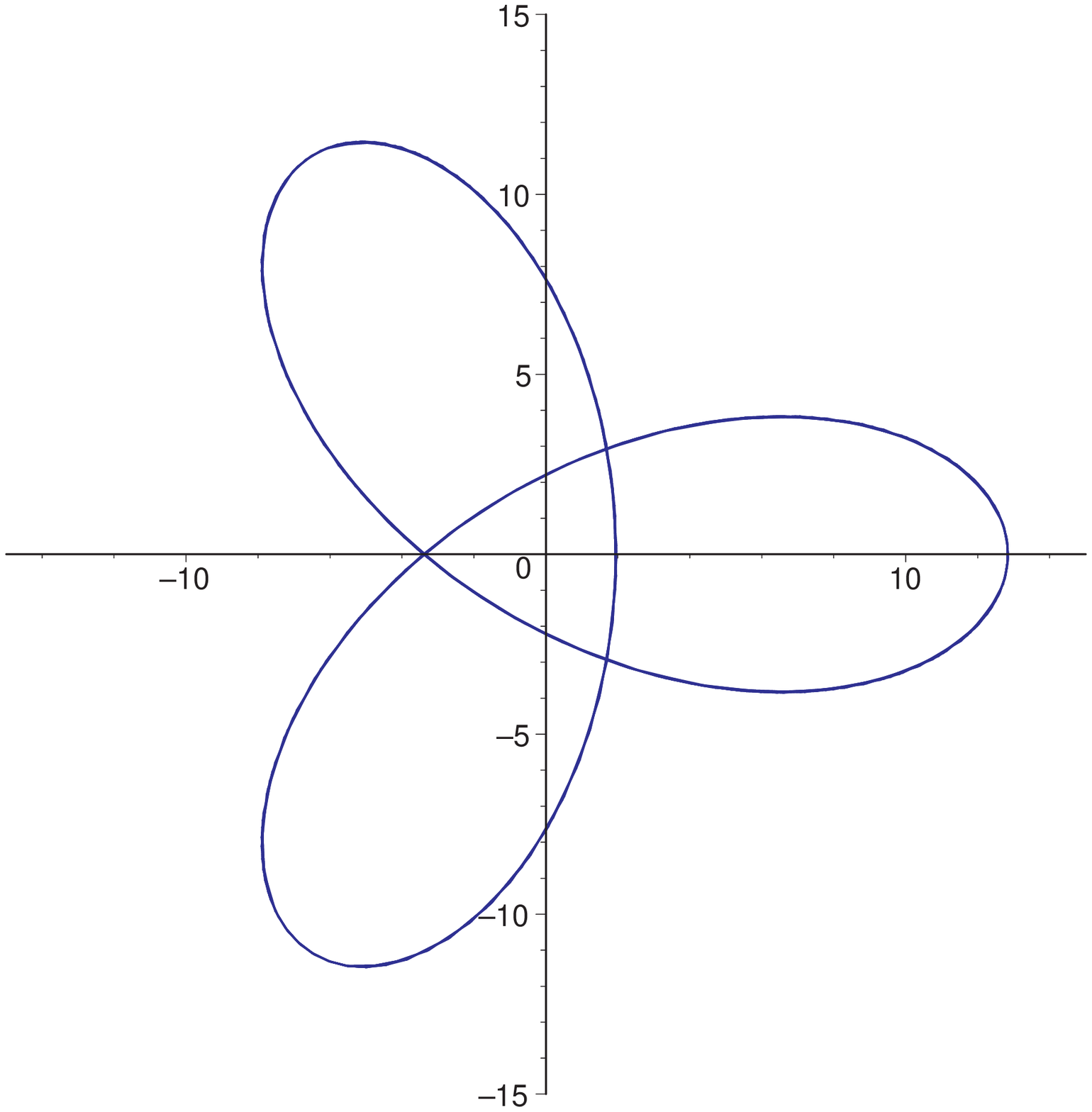}{8.6cm}{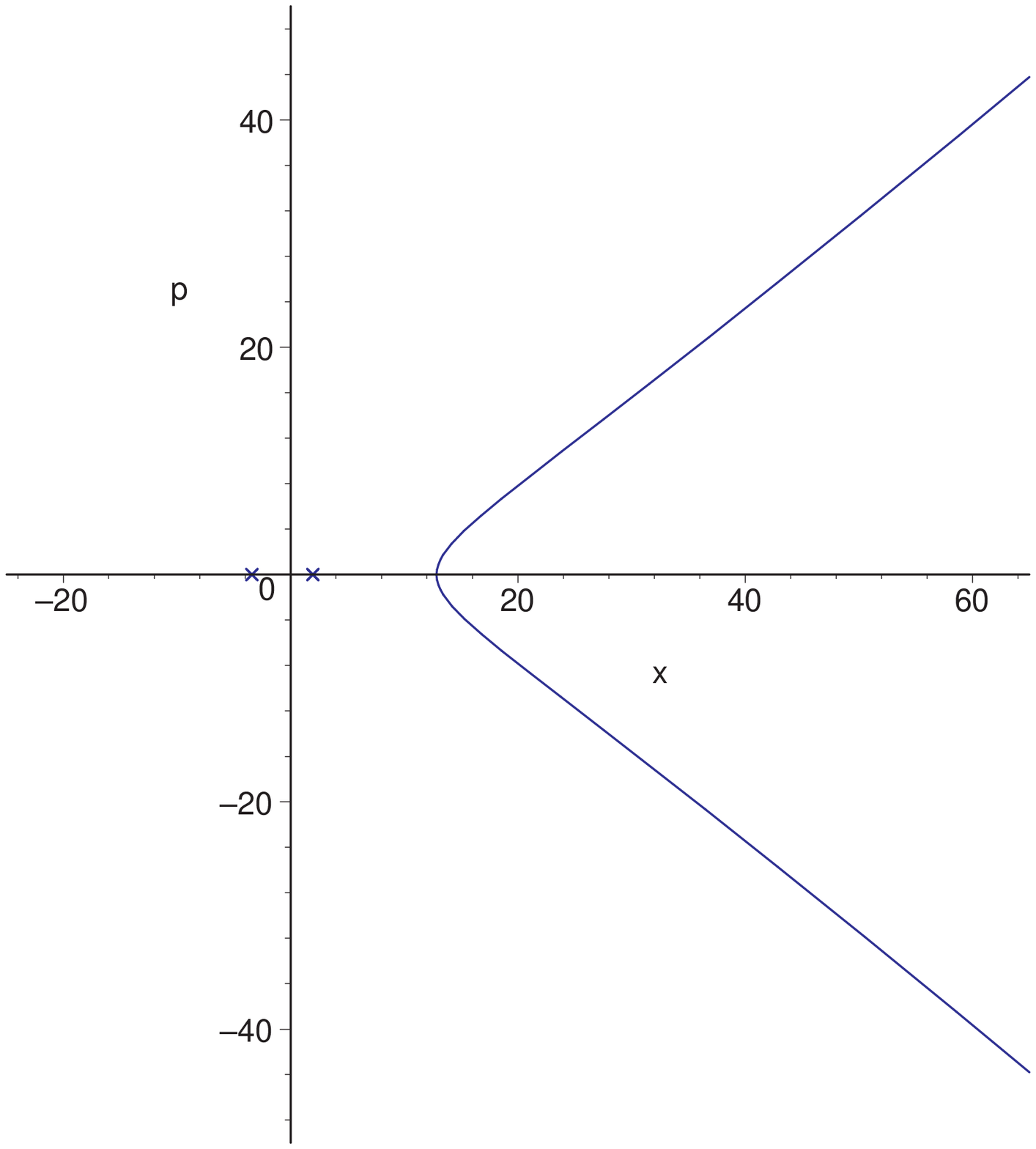}{7.9cm}{traject1}

Let $R=r/s$ with relatively prime integers $r$ and $s$.
From the solution \Ref{xpmo}, it is clear that in this case
nothing changes after shifting the uniformization parameter by $2\pi r$.
This indicates that the complex curve wraps a figure in $\Cb^2$
infinitely many times as in the non-perturbed case. However, now the figure
is more complicated than the hyperboloid \Ref{hyperb}.

To understand its structure, let us consider
a section of the Riemann surface by the plane
\be
\xp=\xm^*.
\label{planez}
\ee
The section is given by some closed curve which coincides with the trajectory
of a fermion in Euclidean time discussed in the previous section.
Several examples of such trajectories for various values of the radius are given in
fig. \ref{traject1}.a and \ref{traject2}.
It has $2r$ intersections with the $p=0$ axis.
Among them two are simple intersections and at $2r-2$ points on the $p=0$ axis
the curve has self-intersections.
They correspond to $r$ non-trivial solutions of the equation \Ref{solz}
and are given in \Ref{points}.
All other solutions can be obtained by either adding $2\pi rn$ or inverting the sign.

The solutions $\phi_{rn} =\pi r n$ are special. They are exact solutions and do not get
$\lambda$ corrections. They can be also written as $\phi_{rn}=\pi R sn=\tilde \phi_{sn}$
what shows that the corresponding instanton contributions
coincide with the T-dual ones.
The special role of these solutions is obvious also from the geometric picture:
they give just the two points on the $p=0$ axis where the curve does not have
self-intersections.

Note that there are also many other self-intersections of the curve.
All of them can be obtained in two ways: either from the self-intersections of the original curve
belonging to the $p=0$ axis by rotating it to the angle $2\pi R n$, or from the curve 
obtained with $-\lambda$ by rotating it to the angle $(2n-1)\pi R$. 
Both transformations are symmetries of the curve.

\twofig{Trajectories in Euclidean time for a) $R=3/4$, b) $R=4/7$.}
{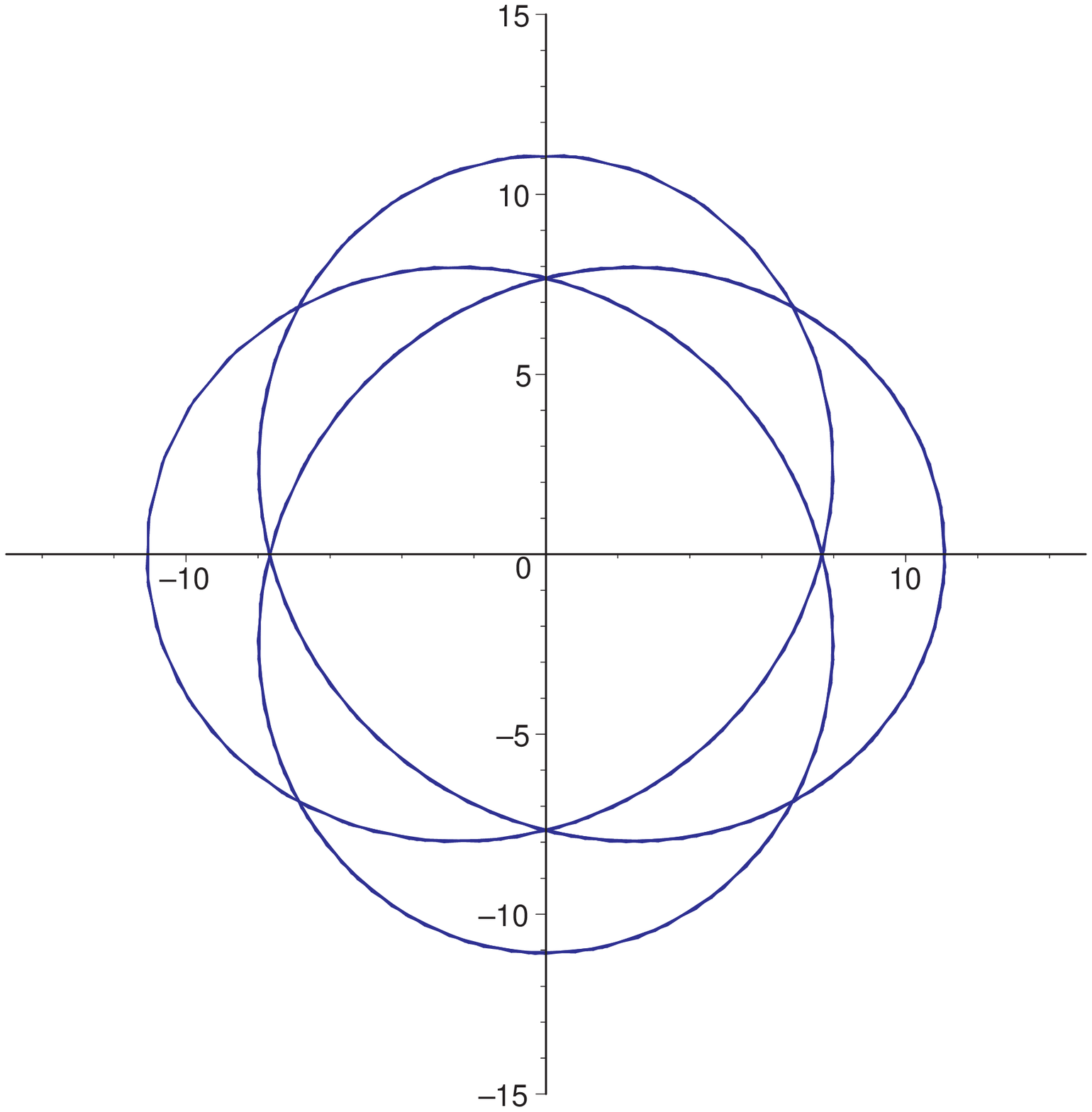}{8.5cm}{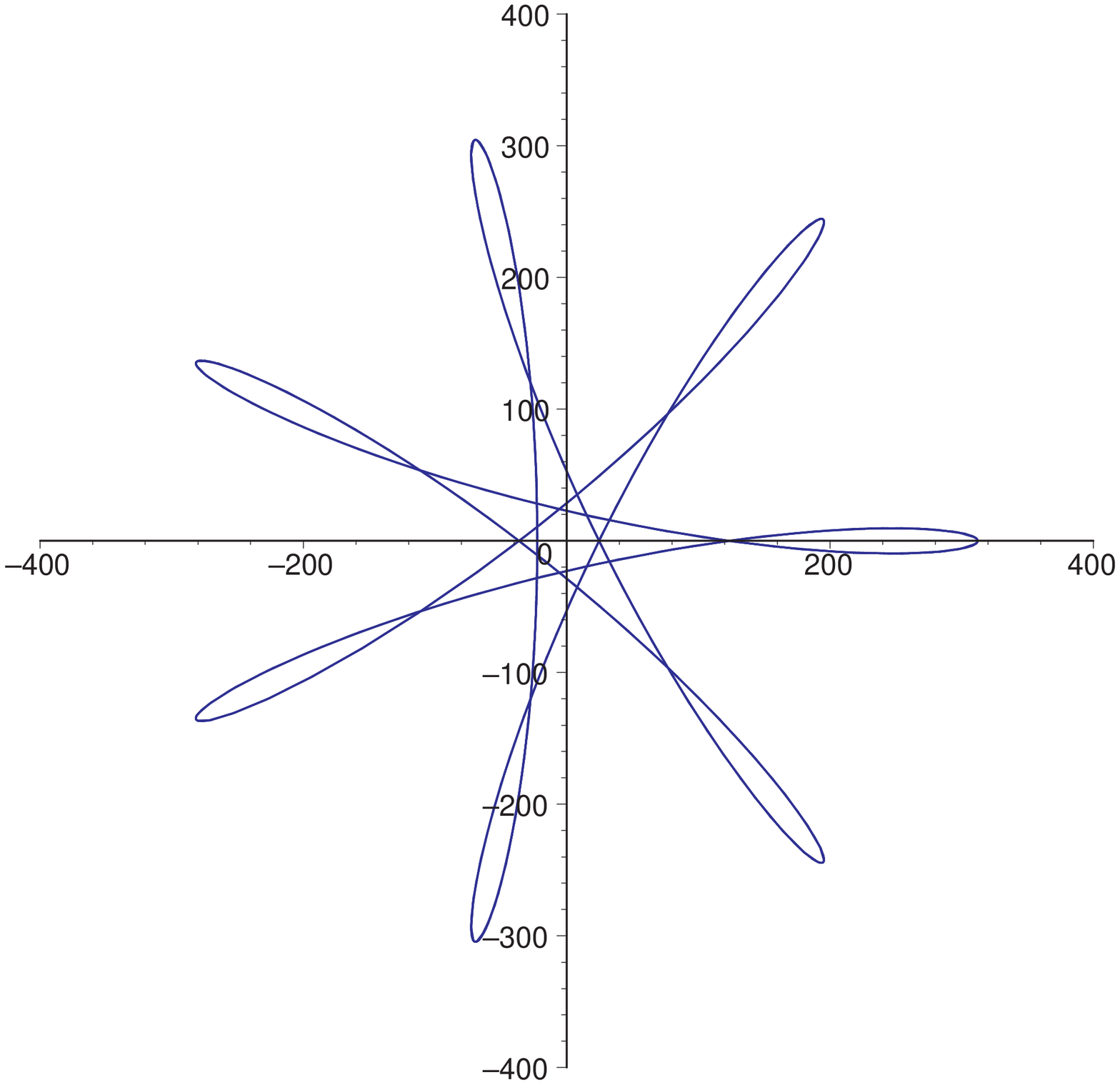}{8.5cm}{traject2}

Another useful characteristic of the Riemann surface is the section by the plane
\be
\xp=\xp^*, \qquad \xm=\xm^*,
\label{planex}
\ee
which corresponds to the real phase space of MQM.
In this case the Riemann surface induces on this plane one hyperbolic like
curve, which coincides with the profile of the perturbed Fermi sea,
and a finite number of points \Ref{points} on the $p=0$ axis
(see fig. \ref{traject1}.b).

It is clear that a similar picture arises on each plane 
\be
\xp=e^{4\pi i R n}\xp^*, \qquad \xm=e^{4\pi i R n}\xm^*,
\label{planerot}
\ee
which is obtained by rotation of the plane \Ref{planex} by $2\pi R n$.
Thus, one has a `Fermi sea' attached to the turning point of each loop 
of the Euclidean trajectory. It is not a real Fermi sea (except the cases when
$2nR\in \Zb$) because its coordinates are complex. It represents the directions
in $\Cb^2$ where the effective Hamiltonian \Ref{fHpert} is unbounded from below.

\subsubsection{Irrational $R$}

In this case the solutions of \Ref{solz}, $\phi_k$, are all irrational
and the corresponding points $x_k$ densely fill the interval
$[-x_0,-b\,x_0]\cup[b\,x_0,x_0]$, where $b={|1-a|\over 1+a}$.
This shows that the degeneracy of the non-perturbed theory is completely
removed and the Riemann surface represents a deformed hyperboloid
wrapping infinitely many times,
but each time along a different trajectory in $\Cb^2$.

This can be illustrated again taking the section by the plane \Ref{planez}.
The section gives a curve densely filling a ring as shown on fig. \ref{trajectir}.
It has infinitely many self-intersections.
All points \Ref{points} correspond to self-intersections since there are
no coincidences between $\phi_k$ and T-dual $\tilde\phi_k$.
The sections by the planes \Ref{planerot} induce the same picture as in the previous case
except that the number of points on the $p=0$ axis is now infinite. 

\lfig{Trajectory in Euclidean time for an irrational value of the radius.}
{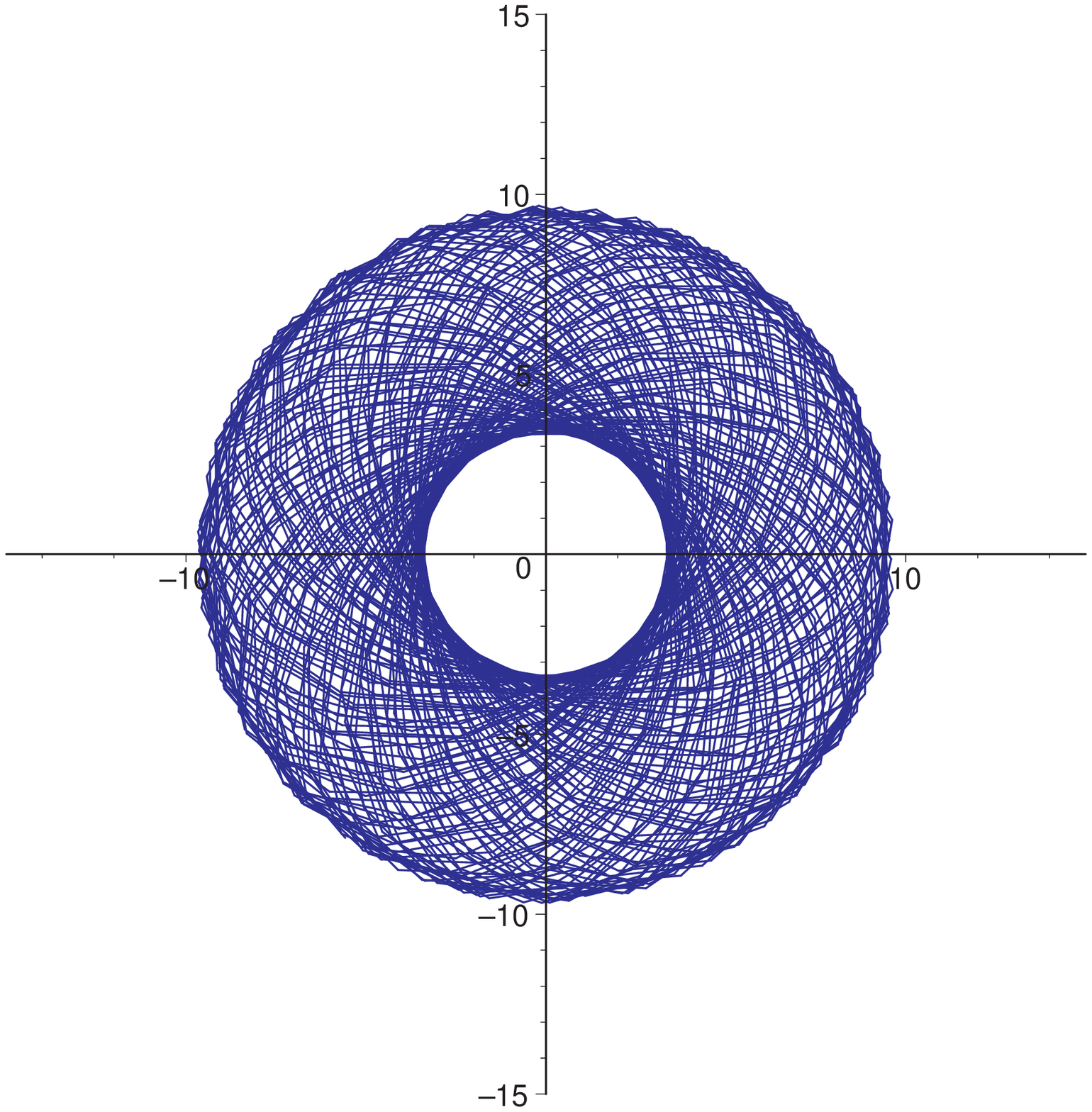}{8.5cm}{trajectir}

\vspace{0.7cm}

We conclude that the Riemann surface $\SMQM$ associated with the solution of the perturbed
MQM has a structure of a deformed hyperboloid wound infinitely many times.
It has singularities which can be interpreted as pinched cycles or self-intersections.
The self-intersections situated on the $p=0$ plane are given by the points \Ref{points}
which, as we saw in the previous subsection, are associated with the instanton effects. 
If one removes the perturbation, the singularities \Ref{points} degenerate
approaching two points $x_k\mathop{\to}\limits_{}(-1)^k \sqrt{\mu}$
and they are distinguished only by the number of windings around
the origin.

The results \Ref{SEucl} and \Ref{SEuclT} imply that
the $k$th instanton contribution can be calculated
as a contour integral of the holomorphic differential $ip\,dx$
around the cycle going through $x_0$ and $x_k$, whereas its T-dual contribution
is given by the same integral along the contour joining two `Fermi seas'.
Therefore, one can associate the two types of instantons with the following trajectories 
of fermions:

1) The non-perturbative contributions \Ref{anz} to the free energy arise when a fermion starts
from the usual Fermi sea (fig. \ref{traject1}.b), moves along the hyperbolic trajectory,
then at the turning point $x_0$ it passes to the Euclidean trajectory, makes a closed loop there
returning to $x_0$, and goes away along the second half of the hyperbolic trajectory.

2) The T-dual non-perturbative contributions, which do not depend on $\lambda$,
appear from fermions which come again from the usual Fermi sea, pass to the Euclidean trajectory
and at the moment $\tau=2\pi k R$ 
pass to the `Fermi sea' attached to the corresponding point of the trajectory. After that they 
go to infinity inside this `complex Fermi sea'.

Thus, to get the former type of corrections the fermion should return to 
the same Fermi sea which it started with, whereas in the latter case it passes to another sea
somewhere in the complex space.

Finally, we make several remarks concerning the constructed complex curve.
First, the complex curve provides a geometric interpretation for the $c=0$ critical limit.
It is known that increasing the Sine--Liouville coupling $\lambda$
one reaches a critical point where
the system behaves as the $c=0$ theory \cite{HSU}.
It was shown that the function $\phi_1(y)$ vanishes in this limit \cite{KAK}.
Thus, the critical behaviour happens when a singularity
of the Riemann surface approaches a turning point, $x_1\to x_0$.
Note that other singularities remain away
from this point what means that the non-perturbative
effects associated with them disappear in the critical limit \cite{SAmn}.

The second remark is that $\SMQM$ is actually the same complex curve which was considered
in \cite{AKKNMM,Boyar}. However, in that papers the curve was investigated
in the coordinates $(\xp^{1/R},\xm^{1/R})$. In these coordinates it looks quite different.
In particular, the section by the plane
\be
\xp^{1/R}=(\xm^{1/R})^*,
\label{planezz}
\ee
which plays the role of the plane \Ref{planez} in the new coordinates,
always induced a closed curve.
Therefore, the singularities, which we analyzed here and which
are related to the non-perturbative effects, were not seen.

Finally, $\SMQM$ should have a direct relation to the ground ring
of the perturbed $c=1$ theory
because it is defined by a deformation of the ground ring equation \cite{Witgr}.
Besides, a similar curve in the $c<1$ case also has such a relation \cite{Seib}.

\section{Complex curve and D-branes in $c=1$ string theory}
\label{CFTsection}

In this section we turn to the CFT description of the $c=1$ string theory.
We show how the complex curve described in the previous section
arises in the CFT framework from the disk partition function
and how it incorporates information about various D-branes.
We start by reviewing some known facts about D-branes in Liouville theory
(which are the main non-trivial part of the $c=1$ branes). A review of
recent developments in Liouville theory can be found in \cite{Nakayama}.

\subsection{Review of D-branes in Liouville theory}

Liouville theory appears when one considers (a matter coupled to)
two-dimensional gravity in the conformal gauge.
It is defined by the following action:
\be
S_L= {1\over 4\pi}\int d^2\sigma \left[ (\p\phi )^2 + Q\hat R\phi
+\mul e^{2b \phi} \right].
\plabel{LIOU}
\ee
The central charge of this CFT is given by
\be
c_L=1+6Q^2
\plabel{cliouv}
\ee
and the parameter $b$ is related to Q via the relation
\be
Q=b+{1\over b}.
\plabel{bQ}
\ee
In general, $b$ and $Q$ are determined by the requirement that the
total central charge
of matter and the Liouville field is equal to $26$.
In the case when matter is represented by a
minimal $(p,q)$ model with the central charge
$c_{p,q}=1-6{(p-q)^2\over pq}$, the relation \Ref{cliouv} implies that
\be
b=\sqrt{p/q}.
\plabel{bminmod}
\ee
The coupling to the $c=1$ matter corresponds to the limit $b\to 1$.
Although we are mostly interested in this limit, usually, quantities found
in Liouville theory have singularities at this point which indicates
that Liouville couplings should be renormalized. Therefore, sometimes we keep
the parameter $b$ arbitrary and take the limit in the final results.

An important class of conformal primaries in Liouville
theory corresponds to the operators
\be
V_\alpha(\phi)=e^{2\alpha\phi}
\plabel{opalph}
\ee
whose scaling dimension is given by $\Delta_\alpha=
\bar\Delta_\alpha=\alpha(Q-\alpha)$.
The Liouville interaction in \Ref{LIOU} is $\delta\CL=\mul V_b$.

The boundary wave functions corresponding to Neumann boundary conditions (FZZ branes)
were constructed in \cite{FZZbrane}.
There is a one parameter family of Neumann boundary conditions which are labeled
by the boundary cosmological constant $\mub$.
It is convenient, however, to introduce a new parameter $\ss$
via the relation
\be
{\mub}=\sqrt{\mul\over \sin(\pi b^2)}\cosh(\pi b \ss) .
\label{fzzsmub}
\ee
Then the Liouville boundary state
is written as
\be
\langle B_\ss| = \int_{0}^{\infty}dP\, \cos
(2\pi P \ss) \Psi(P)\langle P|,
\label{FZZTbs}
\ee
where $\langle P|$ is the Ishibashi state for the Liouville field and
\be
\Psi(P)=\DD \(\pi\mul\gamma(b^2)\)^{-{iP\over b}}
{\Gamma(1+2iPb)\Gamma\left(1+{2iP/b}\right)\over i\pi P}.
\label{FZZTwvfn}
\ee
Here $\gamma(x)={\Gamma(x)\over \Gamma(1-x)}$ and $\DD$ is some numerical constant independent
of all parameters.
The full boundary wave function $\cos(2\pi P\, \ss)\Psi(P)$ includes both
$\ss$- and $P$-dependent parts. It gives one-point correlation functions
of the vertex operators \Ref{opalph} on the disk with the Neumann boundary
conditions defined by $\mub$ \Ref{fzzsmub}.
The label of the vertex operators is related to the momentum $P$ via
$iP=\alpha-Q/2$.

We included the undetermined constant $\DD$ to have {\it exact} coincidence  
of the boundary wave function and the one-point correlators. 
It is difficult to fix it directly from this condition.
(See, for example, \cite{KAK} for discussion of the related question for the Dirichlet
boundary function.) In principle, it can be found from the multi-point correlation functions 
\cite{KostovGR}, but we will leave it non-fixed since it does not influence our results.

The boundary states corresponding to Dirichlet boundary conditions (ZZ branes)
are given by \cite{ZZ}
\be
\langle B_{m,n}|= 2 C \int_{0}^{\infty}dP\,
\sinh (2\pi n P b)\sinh ({2\pi m P/ b})\Psi(P)\langle P|
\label{ZZbs}
\ee
with the same $\Psi(P)$ as in \Ref{FZZTwvfn}
and
\be
C={\DD^{-1} \over  2^{3/4}\sqrt{\pi}}.
\label{Cconst}
\ee
The normalization coefficient was discussed in detail in \cite{KAK}.
We choose it as in that work where it was shown that this normalization
reproduces the matrix model result \Ref{fphi} for the non-perturbative corrections.
An important fact is that
the ZZ state can be obtained as a difference of two FZZ states 
\cite{Hos,MARTINEC,Seib}
\be
\langle B_{m,n}|=C\[\langle B_{\ss(m,n)}|-\langle B_{\ss(m,-n)}|\],
\qquad \ss(m,n)=i\left({m\over b}+n b\right)
\label{zzrel}
\ee
what follows from the comparison of \Ref{FZZTbs} and \Ref{ZZbs}.
The boundary cosmological constant corresponding to $\ss(m,n)$ is
\be
\mub(m,n)=\sqrt{\mul \over\sin(\pi b^2)}\,(-1)^m\cos \pi n b^2,
\label{zzmuB}
\ee
and due to $\mub(m,-n)=\mub(m,n)$ we subtract two states with the same
boundary cosmological constant.

In the $c=1$ limit the distinction between $(m,n)$ and $(n,m)$ branes disappears.
Besides, all $(m,n)$ branes correspond to the same (up to sign) boundary
cosmological constant
\be
\mubc(m,n)= (-1)^{m+n}\sqrt{\muc},
\label{mubc}
\ee
where we introduced the renormalized Liouville couplings
\be
\muc=\mathop{\lim}\limits_{b\to 1} \[\pi(1-b^2)\mul\] ,
\qquad
\mubc=\mathop{\lim}\limits_{b\to 1} \[\pi(1-b^2)\mub\].
\label{mucone}
\ee
This degeneracy corresponds to the degeneracy of singularities of
the complex curve $\SMQM$ of the non-perturbed MQM found
in the previous section. We expect that it is removed by the introduction of a perturbation.

\subsection{Complex curve from FZZ boundary state}
\label{curveFZZ}

Following to \cite{Seib}, we expect that all information about D-branes can be encoded
in the Riemann surface coming from
the partition function of the FZZ boundary state. Since the $c=1$ limit
is little bit subtle and since
we want to trace out also all normalization coefficients,
we start by repeating the calculation of \cite{Seib}.
We will show that even all coefficients fit nicely into the unified picture.

The FZZ partition function can be calculated by integrating the one-point correlator
of the cosmological constant operator\footnote{The minus
sign comes from the minus in front of the action in
the Euclidean path integral.}
\beq
-\left. \p_{\mul} \ZFZZ\right|_{\mub}&=&\langl V_b \rangl_{\ss}
=\cosh\((2b-Q)\pi \ss\)\Psi(i(Q/2-b))\ZD \nonumber \\
&=& 2^{3/4}\DD \, \(\pi\mul\gamma(b^2)\)^{\hf\({1\over b^2}-1\)}
{\Gamma\(b^2\)\Gamma\( 2-1/b^2\) \over (b-1/b)\pi\sqrt{R}}\cosh\((b-1/b)\pi \ss\),
\label{onpcosm}
\eeq
where we used the disk partition function of the compactified scalar field with Dirichlet
boundary conditions (see, for example, \cite{KAK})
\be
\ZD=2^{-1/4}/\sqrt{R}.
\label{Zdiskone}
\ee
The integration gives the following result
\beq
\ZFZZ&=& -2^{7/4}\DD\, \(\pi\mul\gamma(b^2)\)^{\hf\({1\over b^2}+1\)}
{\Gamma\(1-b^2\)\Gamma\( 2-1/b^2\) \over b(1-1/b^4)\pi^2\sqrt{R}}
\nonumber \\
&\times &
\( b^2\cosh\(\pi b\ss\)\cosh\(\pi \ss/b\)-
\sinh\(\pi b\ss\)\sinh\(\pi \ss/b\)\).
\label{Zdisk}
\eeq
As in \cite{Seib} we take the derivative with respect to $\mub$ and get
\be
\left. \p_{\mub} \ZFZZ\right|_{\mul}=-\DD\,{2^{7/4}b\Gamma\( 2-1/b^2\) \over \pi\sqrt{R}}
\, \(\pi\mul\gamma(b^2)\)^{1\over  2 b^2}
\,
\cosh\(\pi \ss/b\).
\label{Zmub}
\ee

In the case of minimal models coupled to gravity
the complex curve describing the ground ring and
giving a geometric interpretation to D-branes arose as the Riemann surface
of the quantity \Ref{Zmub} considered as
a function of $\mub/\sqrt{\mul}$ \cite{Seib}. However, the naive limit $b\to 1$
in \Ref{Zmub} leads to a trivial term proportional to $\mubc$.
In fact, taking into account the renormalization \Ref{mucone}, one observes
that this non-universal leading term is divergent and should be subtracted.
Therefore, we define
\beq
w(\ss)&\equiv& \mathop{\lim}\limits_{b\to 1}\[{1\over \pi(1-b^2)}
\( \p_{\mub} \ZFZZ + 4\DD \, \ZD\,
 {b\Gamma\( 2-1/b^2\)\over \Gamma(1-b^2)}\,\mub \)\]
\nonumber \\
&=& 
-\tDD\sqrt{\muc}
\(\hf \cosh(\pi\ss)\log\muc + \pi\ss \sinh(\pi \ss)\),
\label{redZmub}
\eeq 
where we introduced the normalization coefficient
\be
\tDD={2^{7/4}\over \pi^2\sqrt{R}}\,\DD= {4\over \pi^2}\,\DD \, \ZD.
\label{newC}
\ee
Considered as function of the complex parameter
\be
\mubc=\sqrt{\muc}\cosh(\pi \ss),
\label{mubmuc}
\ee
the quantity \Ref{redZmub} defines
a non-trivial Riemann surface $\SFZZ$.
As in the case of $c<1$ string theories, it provides a geometric interpretation
to all D-branes.

First of all, the FZZ partition function can be considered as a line integral
from some fixed point $\CP\in \SFZZ$
of the one-form $w\, d\mubc$
\be
\ZFZZ=\int_{\CP}^{\mub} w\, d\mubc.
\label{ZFZZint}
\ee
\lfig{Section of the complex curve $\SFZZ$ by the plane $\mubc=\mubc^*$, $w=w^*$
(with $R=1/\sqrt{2}$, $\muc=1$). The singularities correspond to two points \Ref{twopoint}
and ZZ branes are associated with closed cycles passing these singularities.}
{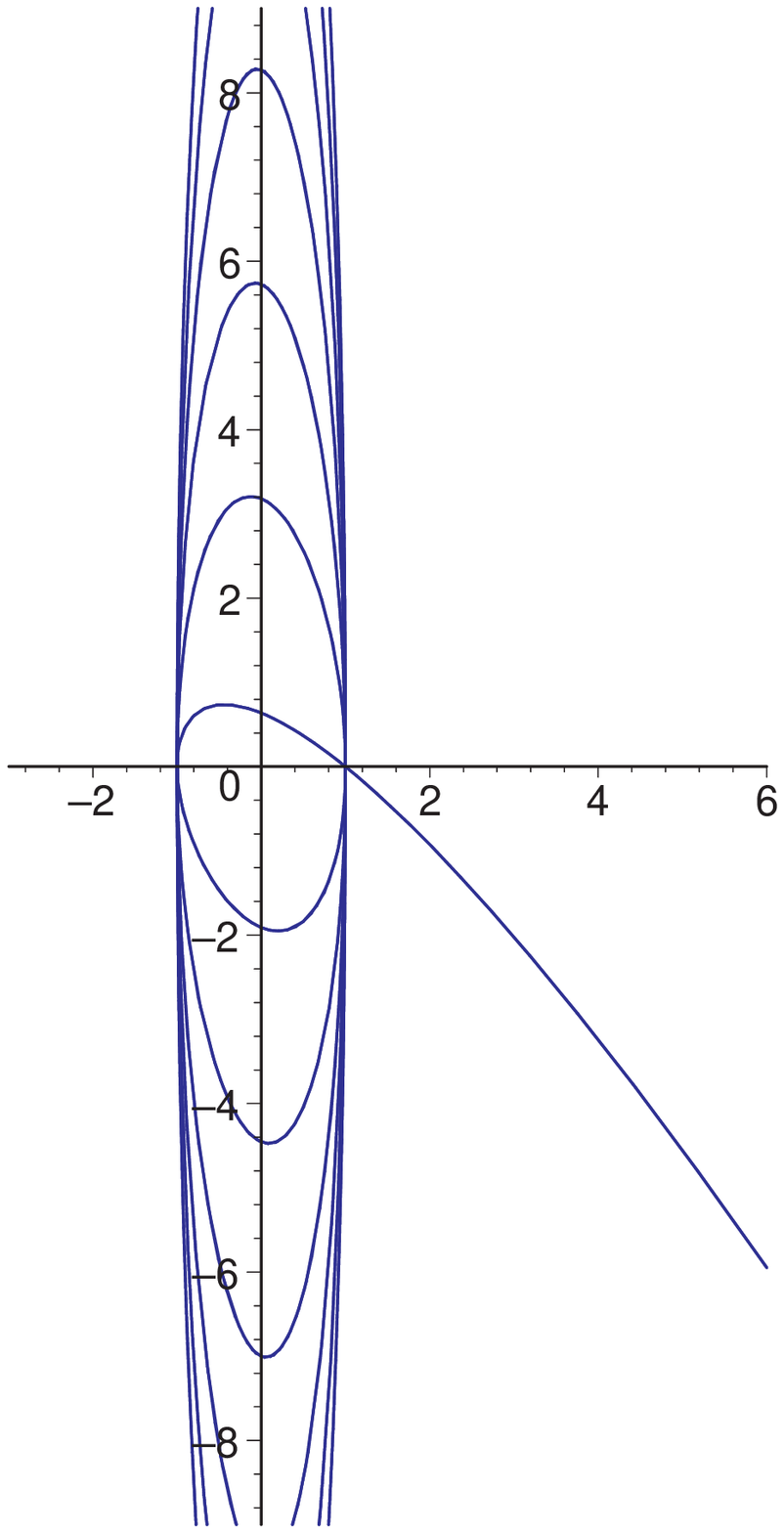}{3.05cm}{fzzcont}
To understand the geometric origin of ZZ branes, we should analyze
the singularities of $\SFZZ$ \cite{Seib}. The singularities correspond to the points
where the complex curve has self-intersections or, in other words, two values
of $\ss$ give the same coordinates $(\mubc,w)$. It is clear that
there is a one-parameter set of such values, $\ss_k=i k$, $k\in \Nb$,
which is the $b\to 1$ limit of $\ss(m,n)$ from \Ref{zzrel}.\footnote{Both $\mubc$ and $w$
are symmetric functions of $\ss$. Therefore, varying $\ss$ over the whole complex plane,
we cover the curve two times and this symmetry does not produce singularities.}
All of them correspond only to two points on the complex curve:
\be
\((-1)^k\sqrt{\muc}\, ,\, {\tDD\over 2}(-1)^{k+1}
\sqrt{\muc}\log\muc\).
\label{twopoint}
\ee
Thus, in this case the singularities are infinitely degenerate and all $(m,n)$ ZZ branes
sit at the points \Ref{twopoint}.
This situation is illustrated on fig. \ref{fzzcont} by a particular section
of the surface $\SFZZ$.
Evaluating the integral of the one-form $w\, d\mubc$
around a cycle on $\SFZZ$
going from $\ss_{k_1}$ to $\ss_{k_2}$ ($k_1$ and $k_2$ should be
of the same parity to have a closed cycle) and taking into account the relation \Ref{zzrel},
one finds
\be
\oint_{k_1 \to k_2} w\, d\mubc=\ZFZZ(\ss_{k_1})-\ZFZZ(\ss_{k_2})=
C^{-1} \ZZZ\({k_1+k_2 \over 2},{k_1-k_2 \over 2}\) .
\label{ZZoint}
\ee
Thus, as in the $c<1$ case, the ZZ branes can be thought as closed cycles running
through singularities of the Riemann surface $\SFZZ$ of the FZZ partition function.
Actually, the relation \Ref{ZZoint} holds also for the boundary states and not only for the
partition functions.

\subsection{Relation between FZZ and MQM complex curves}
\label{relMQMCFT}

However, it is clear that the Riemann surface $\SFZZ$ does not coincide with
the matrix model hyperboloid \Ref{hyperb}.
In this section we show how the matrix model curve $\SMQM$ can be constructed from the FZZ
partition function.

It is clear that the one-point boundary correlator \Ref{redZmub}
is related to the resolvent of the corresponding matrix model.
On the other hand, the hyperboloid of MQM arises from the density function.
Due to this, it is natural to define
\be
\rho= -{1\over 2\pi i }
\(w\(\ss+i\) -w\(\ss-i\)\),
\label{Riempar}
\ee
where $w$ is related to $\ZFZZ$ as in \Ref{redZmub}.
We claim that the function $\rho(\mubc)$ is the same, up to normalization which
will be established soon, as the function
$p(x)$ defined in MQM. As a result, the Riemann surface defined by $\rho(\mubc)$,
we will call it $\SCFT_1$, coincides with the complex curve arising in the matrix model
description of the $c=1$ string theory:
$\SCFT_1=\SMQM$.
This statement is supposed to be true not only in the original $c=1$ string theory
but also in the perturbed one.
We will verify it in the first two orders in the perturbative expansion in
the Sine-Liouville coupling $\lambda$.

Since the curve $\SMQM$ describes the ground ring of the $c=1$ string theory,
\Ref{Riempar} gives a relation between the bulk ground ring and the disk partition function.
The origin of this relation is not evident. We believe that it can be better understood
from the results found in \cite{KostovSolv,KostovPS,KostovGR},
where a similar relation appeared between the density and the resolvent 
of a matrix chain. The latter can be thought as a discretization of MQM
at distance $\pi/2$. However, an interpretation of this relation directly from the CFT point of view
is still missing.

Let us check that our identification is correct comparing
the complex curves coming from the matrix model and from D-branes
up to the first order in the perturbative expansion in $\lambda$.
The matrix model curve $\SMQM$ is defined by \Ref{xpmo} and in the coordinates $(x,p)$
takes the form
\beq
x(\tau)&=&\sqrt{2\mu}\( \cos\tau +{\lambda\over R}\mu^{{1\over 2R}-1}
\cos\(\(1-\oR\)\tau\)\)+O(\lambda^2),
\label{xtau}
\\
p(\tau)&=&-i\sqrt{2\mu}\( \sin\tau +{\lambda\over R}\mu^{{1\over 2R}-1}
\sin\(\(1-\oR\)\tau\)\)+O(\lambda^2).
\label{ptau}
\eeq
Solving the first equation with respect to $\tau$, one finds
\be
\tau(x)=\tau_0(x)+ {\lambda\over R}\, \mu^{{1\over 2R}-1}\,
{\cos\(\(1-\oR\)\tau_0(x)\)\over \sin \tau_0 (x)} +O(\lambda^2),
\label{soltau}
\ee
where
\be
x=\sqrt{2\mu}\,\cos\tau_0(x).
\label{tauzero}
\ee
Finally, the substitution of \Ref{soltau} into \Ref{ptau} gives
\be
p(\tau_0)=-i\sqrt{2\mu}\(\sin\tau_0+{\lambda\over R}\, \mu^{{1\over 2R}-1}\,
{\cos{\tau_0\over R}\over \sin \tau_0 }  \)+O(\lambda^2).
\label{pofx}
\ee

Now we consider the function $\rho(\ss)$ in the theory with
the Sine--Liouville perturbation.
The leading order in $\lambda$ is given by \Ref{Riempar}
with $w(\ss)$ from \Ref{redZmub}.
The result is
\be
\rho_0(\ss)=-\tDD\sqrt{\muc}\sinh(\pi\ss).
\label{ylead}
\ee
The next order term is
\be
\rho_1=\mathop{\lim}\limits_{b\to 1}
\({\laml \langl \cos\(X/R\)\rangl_{\rm Dir}\over 2\pi i}
\[\p_{\mub}\langl V_{b-{1\over 2R}} \rangl_{\ss+i/b} -
\p_{\mub}\langl V_{b-{1\over 2R}} \rangl_{\ss-i/b} \over \pi(1-b^2)\]\) .
\label{ysublead}
\ee
The one-point function of $\cos\(x/R\)$ on the disk with Dirichlet boundary conditions
is the same as \Ref{Zdiskone}
\be
\langl \cos\(X/R\)\rangl_{\rm Dir}=2^{-1/4}/\sqrt{R}.
\label{Zdiskcos}
\ee
From the explicit form of the boundary wave function \Ref{FZZTwvfn}, one finds
\be
\rho_1(\ss)=\laml \(\sqrt{\muc}\)^{\oR-1}
{2^{3/4}\DD\,  \Gamma^2\(1-\oR\)\sin\(\pi\over R\)\over  \pi^2\sqrt{R} }
{\cosh\({\pi\ss\over R}\)\over \sinh(\pi \ss)}.
\label{yone}
\ee
Taking together \Ref{ylead} and \Ref{yone}, one obtains
\be
\rho(\ss)=-\tDD\sqrt{\muc}
\(\sinh(\pi\ss) - {\pi \over 2}\gamma\(1-\oR\){\laml} \muc^{{1\over 2R}-1} \,
{\cosh\({\pi\ss\over R}\)\over \sinh(\pi \ss)} \)
+O(\laml^2).
\label{yfull}
\ee

A direct comparison of \Ref{yfull} and \Ref{pofx} is not straightforward because
the coefficients can depend on a relative normalization of the CFT and matrix model
couplings. This relative normalization can be fixed considering bulk correlators on
a sphere.
This is done in Appendix \ref{A}. The result \Ref{relCFTMQM} together
with \Ref{mubmuc} and \Ref{tauzero} shows that up to a coefficient
the function \Ref{yfull} is exactly the same as \Ref{pofx}.
Thus, we proved that the complex curves $\SMQM$ and $\SCFT_1$ arising
in the perturbed MQM and from the FZZ partition
function of Sine--Liouville theory, respectively, coincide
in the first order in the perturbation coupling. We suggest that
the identification is actually valid to all orders in $\lambda$.

From this result
we also see that the Liouville boundary parameter $\ss$ is
identified with the parameter $\tau_0$ \Ref{tauzero}, which in the leading order
coincides with Euclidean time of the perturbed MQM,
\be
i\tau_0=\pi\ss.
\label{identpar}
\ee
However, our results imply that in the perturbed theory
it is more natural to introduce a $\lambda$-dependent parameterization
of the boundary cosmological constant coming from
the full $\tau$. In particular, it is $\tau$, and not $\ss$, that is the uniformization
parameter of the complex curve $\SCFT_1$.

Let us fix also the relative normalization of $\rho(\mubc)$ and $p(x)$.
First of all, using the relation between $\mul$
and $\mu$ and comparing \Ref{mubmuc} with \Ref{tauzero},
we can write a relation between $\mubc$ and the fermionic coordinate $x$
\be
{\mubc\over x}= {\(\pi^3 R\)^{1/4}\over \sqrt{2}}.
\label{mubx}
\ee
Then the relation between $\rho$ \Ref{yfull}
and $p$ \Ref{pofx} is
\be
\CC\equiv {p\over \rho}={\pi^{5/4}R^{1/4}\over 2^{5/4}\DD}
=\pi C{\mubc\over x}
\label{CCC}
\ee
with the coefficient $C$ from \Ref{Cconst}.
This allows to write \Ref{Riempar} in the following form
\be
p(\tau_0)=-{C \over 2i }
\[\p_{x} \ZFZZ\(i \({\tau_0 / \pi} + 1\)\) -\p_{x} \ZFZZ\(i \({\tau_0 / \pi} - 1\)\)\],
\label{Riem}
\ee
where we used the identifications \Ref{mubx} and \Ref{identpar}.
This equation gives the precise relation of
the complex curve of MQM, describing, in particular,
the profile of the Fermi sea, to the disk partition function
of the FZZ boundary state in the (perturbed) $c=1$ string theory.

\subsection{Complex curve and ZZ branes}

In section \ref{curveFZZ} it was shown that the $(m,n)$ ZZ branes are associated with
the singularities of the Riemann surface $\SFZZ$. What happens with these branes when we pass
from $\SFZZ$ to $\SCFT_1$?
We expect that the branes are again associated with the singularities of the complex curve.
The singularities were analyzed in section \ref{compcurv} and, indeed, they were shown
to give rise to instanton effects.
However, there is only a one-parameter set of the non-perturbative effects,
which was argued to correspond to the $(k,1)$ or $(1,k)$ branes \cite{SAmn}.
Thus, all other $(m,n)$ branes somehow disappear from our description.
Let us study this in more detail.

First, we start with the non-perturbed case. Then the singularities of $\SFZZ$
correspond to the one-parameter set of $\ss_k=ik$, $k\in \Nb$.
The two-parameter set of ZZ branes
appears because each brane is associated with a pair of values of the parameter $\ss$
giving the same point on the curve (they define a closed cycle).
Since all $k$ of the same parity define the same points \Ref{twopoint}, one obtains
that any pair $\ss(m,\pm n)=\ss_{m\pm n}$ defines a ZZ brane.
The same story happens for the non-perturbed complex curve $\SCFT_1$ defined by \Ref{Riempar}.
However, in this case a closed cycle corresponding to the pair $\ss(m,\pm n)$ gives
a difference of $(m+1,n)$ and $(m-1,n)$ branes.

However, when one includes the Sine--Liouville perturbation, the singularities corresponding
to $\ss_k$ move away from their initial positions \Ref{twopoint}
and there is still only a one-parameter set
of the singular points on the curve $\SCFT_1$.\footnote{We consider
only singularities which originate from
\Ref{twopoint}. As it was mentioned in section \ref{compcurv}, there are many
other singularities which all can be obtained by a certain rotation of the complex curve.
They appear when we pass from $\SFZZ$ to $\SCFT_1$ and will be discussed later.}
Indeed, to have a singularity,
one should find a pair of values of $\ss$, say $\ss'$ and $\ss''$,
which give the same point on $\SCFT_1$,
{\it i.e.}, the same boundary cosmological constant $\mubc$ and function $\rho$.
From \Ref{mubmuc} one finds that $\ss''=\pm \ss' +2 i k$. With a non-vanishing $\lambda$
(and generic $R$), $k\ne 0$ can not be realized because the second term in \Ref{yfull}
will be for sure different. Hence, we remain with the only possibility $\ss''=-\ss'$.
Taking into account that even the perturbed $\rho(\ss)$ is antisymmetric, or, equivalently,
the boundary correlator $w(\ss)$ is symmetric, all singularities are given by
equation
\be
\rho(\ss)=0 \Leftrightarrow p(\tau)=0.
\label{eqsing}
\ee
It is easy to check that all solutions of the latter are given by $\phi_k$ from \Ref{solz}
and correspond to the singular points \Ref{points}.

Thus, only a one-parameter set of singularities survives the perturbation.
All of them lie on the $p=0$ axis and flow with $\lambda$ as predicted by
the $\lambda$-dependence of $\phi_k$ \Ref{corr}.
In section \ref{leadinst} it was proven that they are associated with the non-perturbative
effects which, as it was shown in \cite{KAK,SAmn} in the first order in $\lambda$,
arise from the $(k,1)$ branes. (Since in the non-perturbed $c=1$ theory
$(m,n)$ branes are identical to $(n,m)$ branes, we can choose to call the branes
appearing in the perturbed MQM as $(k,1)$.) There are no singularities on $\SCFT_1$
which correspond to $(m,n)$ branes with $n>1$.

Due to this identification one can write a generalization of the relation between
the FZZ and ZZ boundary states \Ref{zzrel} to the case of Sine--Liouville theory.
Since $\SCFT_1=\SMQM$ we can work with the coordinates $(x,p)$ instead of $(\mubc,\rho)$.
Let us pick up a singularity of the Riemann surface $\SCFT_1$ at the point
$(x_k,0)$ \Ref{points}. It corresponds to two values of the uniformization parameter
$\tau=\pm \phi_k$. However, since the relation between the complex curves $\SFZZ$ and $\SCFT_1$
is defined in terms of the parameter $\tau_0$, one should obtain the two values of this
parameter, $\pm \tauk$, corresponding to the position of the singularity.
This can be done by comparing \Ref{points} and \Ref{tauzero} and gives the following
explicit result
\be
\tauk=\pi k+
\arccos\[ y^{-1/2} e^{-{1\over 2R}X(y)}\( \cos\phi_{k}+a(y)\cos\({1-R\over R} \phi_{k}\)\)\].
\label{taumnphi}
\ee
If one wants to write an expansion in the Sine--Liouville coupling, it is enough to substitute
the corresponding expansion of $\phi_k$ \Ref{corr}. Note that near the singularity
the expansion \Ref{soltau}, as well as \Ref{pofx}, breaks down and it cannot be used.
The reason is that all terms turn out to be of the same order $O(\sqrt{\lambda})$.
Fortunately, in this case we have an expression in terms of the uniformization
parameter $\tau$, which allows to find the exact position of the singularity.

Now we take a closed contour $\gamma_{k,1}$ on $\SCFT_1$ starting at $x_k$, 
passing through $x_0$ and returning to $x_k$.
(In terms of the uniformization parameter,
it represents the interval $[-\phi_k,\phi_k]$.)
Then, according to \Ref{SEucl} and \cite{KAK,SAmn},
one can write
\be
\ZZZ(k,1)=i \oint\limits_{\gamma_{k,1}}p\, dx .
\label{ZpZZ}
\ee
On the other hand, using the identification \Ref{Riem}
and the symmetry of the (perturbed) FZZ partition function,
$\ZFZZ(\ss)=\ZFZZ(-\ss)$, one obtains\footnote{To obtain this result, one should take
into account that
\be
\int \p_x \ZFZZ\(i\(\tau_0/\pi+ n\)\) dx (\tau_0) =
(-1)^n \int \p_x \ZFZZ\(i\(\tau_0/\pi+ n\)\) dx (\tau_0+\pi n).
\label{propint}
\ee
}
\be
\ZZZ(k,1)=C\[ \ZFZZ\(i\({\tauk / \pi} + 1\)\) -\ZFZZ\(i\({\tauk / \pi} - 1\)\)\].
\label{ZZint}
\ee
This equation provides a geometric interpretation for the relation \Ref{zzrel}
in terms of the Riemann surface $\SCFT_1$
and gives its generalization to the theory
perturbed by a tachyon potential.

The vanishing of $\rho$ at singularities means that all of them are also singularities
of the complex curve $\SFZZ$ associated with the FZZ partition function.
The result \Ref{ZZint} gives the flow with $\lambda$ of the positions of the $(k,1)$
ZZ branes on $\SFZZ$.
In terms of the boundary cosmological constant their positions are given by
\Ref{points} through the relation \Ref{mubx}.
The two values of the parameter $\ss$, which correspond to the singularity, are
\be
\ss(k,\pm 1)= i \({\tauk \over \pi} \pm 1\).
\label{sigmak}
\ee

\subsection{Complex curves for $(m,n)$ branes}

Although the complex curve $\SCFT_1$ describes only the $(k,1)$ branes,
the initial Riemann surface $\SFZZ$ was able to incorporate all $(m,n)$ branes.
Here we show how they can be obtained in a similar way to the $(k,1)$ branes.

Regarding \Ref{Riempar} and \Ref{sigmak}, it is natural to define
\be
\rho_{(n)}= {(-1)^n\over 2\pi i }
\(w\(\ss+in\) -w\(\ss-in\)\).
\label{Riemparn}
\ee
An important fact is that all these functions can be expressed through $\rho$ \Ref{Riempar}
\be
\rho_{(n)}(\ss)=(-1)^{n-1}\sum_{k=1}^n \rho(\ss+i(n-2k+1)).
\label{ZZnZZ}
\ee
Therefore, using the identification \Ref{CCC} of $\rho$ with the MQM momentum $p$
and the explicit expression of the latter
following from \Ref{xpmo}, in principle, it is possible to find
$\rho_{(n)}$ to all orders in the Sine--Liouville coupling $\lambda$.
However, since the relation between the parameters $\tau$, which is used to write
$p$, and $\tau_0$, which is directly related to $\ss$,
can be found only as a series in $\lambda$, we cannot write a closed
expression for $\rho_{(n)}$. Nevertheless, it is easy to find it in the first order in
$\lambda$. From \Ref{yfull} one finds
\be
\rho_{(n)}(\ss)=-\tDD\sqrt{\muc}
\(n\, \sinh(\pi\ss) - {\pi \over 2}\gamma\(1-\oR\){\laml }\muc^{{1\over 2R}-1} \,
 {\sin\({\pi n\over R}\)\over \sin {\pi \over R}}
{\cosh\({\pi\ss\over R}\)\over \sinh(\pi \ss)} \)
+O(\laml^2).
\label{rhonmom}
\ee
In fact, in this order this expression follows also from the one-point correlator on the disk
with $iP={1\over 2R}-1$ and with the FZZ boundary condition.
But if we are interested in higher orders, they are related to multi-point correlators
which are not known in the CFT formulation.
On the other hand, as we just mentioned, the full expansion of $\rho_{(n)}(\ss)$ 
can be obtained from $p(x)$.
Thus, MQM provides a set of predictions for the multi-point correlators with the FZZ
boundary conditions. In Appendix \ref{B} we discuss the second order in $\lambda$
which predicts the two-point correlator of the Sine--Liouville operator.

Each of the functions $\rho_{(n)}(\mubc)$ defines a Riemann surface $\SCFT_n$.
Let us discuss the singularities of $\SCFT_n$ which should be associated to D-branes
of the $c=1$ string theory.
Repeating the reasoning of the previous section, again one concludes that
all singularities are given by solutions of $\rho_{(n)}(\ss)=0$.
We will call the corresponding values of the parameter $\ss$ by $\ss_{m,n}$
where the index $m$ labels different solutions for a given $n$.
It is clear that 
\be
\ss_{m,n}=im+O(\sqrt{\lambda}).
\label{sigmammn}
\ee
However, we cannot find even the first $\lambda$ correction
because, as it was explained after eq. \Ref{taumnphi}, the perturbative expansion breaks down
near the singularities. Due to this, even the first correction requires the knowledge
of all terms in the expansion of $\rho_{(n)}$. 

In fact, even the existence of the singularities is not guaranteed. 
We expect that if a singularity exists it should be possible to find near it a complex coordinate
in terms of which both $\mubc$ and $\rho_{(n)}$ have a well defined expansion in the Sine--Liouville 
coupling. 
It is possible to find such a parameter in several first orders in $\lambda$ but 
at some point one gets an obstruction. In the case of $n=1$ this obstruction is absent and the 
parameter coincides with the Euclidean MQM time $\tau$.
More details about this can be found in Appendix \ref{C}. 
Thus, it is possible that the perturbation removes all singularities of $\SCFT_n$, $n>1$,
except the trivial one with $m=0$.

Nevertheless, if the singularities on $\SCFT_n$ do exist,
already the simple fact \Ref{sigmamn} is sufficient to show that they would be 
related to the $(m,n)$ ZZ branes. 
Indeed, as usual, each singularity gives rise to a closed cycle.
We choose $\gamma_{m,n}$ as the image of the interval $[-\ss_{m,n},\ss_{m,n}]$
on the plane of the parameter $\ss$. 
Then the explicit expression \Ref{rhonmom} together with \Ref{sigmammn}
allows to find
\be
\pi i \,C\oint\limits_{\gamma_{m,n}}\rho_{(n)}\, d\mubc =
-{2\over\sqrt{\pi R} }\,
\muc\( mn+ {R\over \pi}\Gamma^2\(1-\oR\)\laml\muc^{{1\over 2R}-1}
{\sin\({\pi m\over R}\)\sin\({\pi n\over R}\)} \)+o(\laml).
\label{fmn}
\ee
It is not difficult to check that this result in the two leading orders in $\lambda$
coincides with the partition function $\ZZZ(m,n)$ of the $(m,n)$ ZZ boundary state \Ref{ZZbs}.
As in all previous cases, we suppose that the coincidence remains to be true to all orders.
Thus, the $m$th singularity on $\SCFT_n$ is indeed associated with the $(m,n)$ brane. 

On the other hand, from the definition \Ref{Riemparn}
and taking into account \Ref{propint}, one obtains
\be
\ZZZ(m,n)=\pi i\, C \oint\limits_{\gamma_{m,n}}\rho_{(n)}\, d\mubc =
 C\[\ZFZZ(\ss_{m,n}+i n) -\ZFZZ(\ss_{m,n}-i n)\].
\label{ZpZZmn}
\ee
This relation provides a generalization of \Ref{zzrel} to the perturbed case for all branes.
Besides, the result \Ref{ZpZZmn} gives also the positions of the $(m,n)$ branes 
on the Riemann surface $\SFZZ$.
They correspond to the following values of the parameter $\ss$
\be
\ss(m,n)= \ss_{m,n}\pm i n.
\label{sigmamn}
\ee
Let us repeat that these results were obtained assuming that the singularities 
are present, whereas
the analysis of Appendix \ref{C}
shows that they most likely disappear after the perturbation. 
In the non-perturbed case they are not independent and completely determined by the singularities
of $\SCFT_1$, {\it i.e.}, by the $(m,1)$ branes.

\subsection{Rotation symmetry of the complex curve}

Let us return to the complex curve $\SCFT_1$. As we saw in section \ref{compcurv},
the curve has singularities not only on the axis $p=0$, which we discussed so far.
Since we expect that all singularities are related to some non-perturbative effects,
it is interesting to understand the origin of these additional ones.

Let us recall the fact, which we already mentioned,
that all singularities can be obtained from the singularities 
belonging to the $p=0$ axis by a certain rotation of the curve.
There are two cases. In the first one, one should rotate the curve by angle $2\pi R n$.
In the second case, one should consider the curve for the Sine--Liouville coupling 
of a different sign and rotate it by $(2n-1)\pi R$.
The former transformation is a symmetry of $\SCFT_1$ and the latter is a global rotation
which does not change the form of the original curve.

Due to this, closed cycles going through one of these additional singularities 
can be represented as a combination of the cycles $\gamma_{k,1}$ for the transformed curve
and the corresponding boundary state is given by a linear combination of the states 
we considered before.
Moreover, it is more natural to think about them as $(k,1)$ branes for the `rotated' theory.
The `rotated' theory can be seen as a theory associated with a different section 
of $\SCFT_1$, which produces one of the `Fermi seas' we met in section \ref{compcurv}.

In fact, since the curve is invariant under rotations by $\alpha_n=2\pi R n$,
there appears an interesting duality. When this symmetry is formulated in the CFT terms
it means that the theory with
\beq
\mubc' & =&  \mubc\cos \alpha_n - i\tDD^{-1}\rho\,\sin\alpha_n ,
 \label{rotub} \\
\rho' & =& -i \tDD\mubc\sin\alpha_n  +\rho\,\cos\alpha_n 
 \label{rotrho}
 \eeq
has the same instanton effects as the original one.
(Note that this symmetry does not concern the FZZ branes.)
This is a symmetry of the full theory with a non-vanishing Sine--Liouville coupling.
It is clear that it is related to the discrete symmetry of the Sine--Liouville action 
\Ref{conepert} under translations of the matter field $X\to X+2\pi n R$. 
But now it is realized in the open string sector and it may have some non-trivial consequences.
It would be interesting to understand whether it is trivial or indeed can provide 
something new.

\section{Discussion}

In this paper we found a geometric description in terms of a Riemann surface
for non-perturbative effects in the $c=1$ string theory with the Sine--Liouville 
potential which generalizes the corresponding results for $c<1$ theories \cite{Seib,KK}.
The main conclusions are essentially the same as in \cite{Seib}:

1) The Riemann surface is provided by the partition function of the FZZ boundary state.

2) A line integral of a one-form on the Riemann surface with a fixed reference point gives
the FZZ partition function, whereas the integrals around closed contours passing through 
pinched cycles reproduce the partition functions of the ZZ branes.

3) The partition functions  of the ZZ branes can be obtained as a difference of the FZZ
partition functions taken at two values of a complex parameter corresponding to the same
point (singularity) on the curve \Ref{ZZint}.

However, contrary to \cite{Seib}, the Riemann surface arising here is not the same as 
the one appearing in the matrix model and describing the ground ring.
The exact relation between them is given by \Ref{Riempar} or \Ref{Riem}.
Note that according to this identification the boundary cosmological constant 
is mapped to the matrix eigenvalue which is identical to the fermionic coordinate.

The origin of this identification of the complex curve describing a closed string background
with a curve of open string theory is not quite clear for us.
It can be justified from the matrix model point of view following \cite{KostovGR},
but its direct CFT interpretation is lacking. 
In fact, it implies a non-trivial relation between the ground ring of bulk operators
and D-branes of the $c=1$ string theory. 
This relation is an example of the open/closed string duality and it would be quite 
interesting to understand it in detail.

All our results are supposed to be true also for a finite perturbation by tachyons 
or windings. In this paper we restricted ourselves to the Sine--Liouville potential
corresponding to the simplest time-dependent tachyon condensate. 
Since the corresponding matrix model
is exactly solvable, our results provide some new predictions for the CFT correlation functions
of open strings.   
As a concrete example, we calculated a two-point bulk
correlator on the disk with the FZZ boundary conditions \Ref{predtwo}. 
The main interesting feature of this correlator is the presence of pole singularities 
at the points corresponding to the $(m,n)$ ZZ branes. 
On can draw an analogy with a propagator in QFT which has poles at the values of momentum
corresponding to physical states. Then according to this picture the $(m,n)$ branes would 
play the role of such `physical' states.

Unfortunately, we were not able to make a definitive conclusion about the fate of the $(m,n)$
branes with $n>1$ in the $c=1$ string theory.
Our analysis implies that they may disappear after the perturbation. However, to make
a rigorous statement, one should know how to sum up the perturbation series 
for the FZZ partition function, which we could not do.

Since the FZZ partition function is related to the resolvent of MQM, one could think that 
the integrability allows to find it exactly. However, the integrability is present 
in the chiral coordinates, whereas we need the resolvent which is a function of the original
coordinate $x$. It is not clear how to translate the results from the light cone
formulation to this representation. Actually, this reflects the usual problem.
We had to work in the $x$-representation to establish connection with the CFT results.
At the same time, MQM provides another set of variables which is more natural, but
its interpretation in the CFT terms is obscure.

Returning to the question about the $(m,n)$ branes, 
let us note that 
it is natural if the $(m,n)$ branes do disappear because only the Riemann surface $\SCFT_1$
has an interpretation in the matrix model.
All other surfaces $\SCFT_n$ seems to be auxiliary. 
Besides, there is no place in MQM for the non-perturbative corrections
corresponding to all $(m,n)$ branes.

Another hint to this issue is provided by the above mentioned two-point correlator.
It is natural to expect that not only this, but also all multi-point correlation functions
diverge at the points of the $(m,n)$ branes. 
Therefore, there is a question whether the whole series defining the perturbed disk 
partition function diverges or not at these points.
The finiteness of the results for the $(m,1)$
branes indicate that in this case, nevertheless, it is possible to sum up 
the perturbation series to a well defined function. 
However, for general $m$ and $n$ it may be impossible what indicates the disappearance of the
$(m,n)$ branes.

Finally, we mention that it would be natural to generalize the results obtained in this paper
and in \cite{Seib} to the case of the supersymmetric $\hat c=1$ theory \cite{DKKMMS}. 
The only restriction which may appear is
to find the corresponding complex curve for a finite tachyon perturbation 
because no explicit solution is known for MQM in the two-side inverse oscillator potential 
with the Fermi sea perturbed from both sides.

\section*{Acknowledgements}

The author is grateful to Vladimir Kazakov, 
B\'en\'edicte Ponsot, Marika Taylor, Stefan Vandoren 
and especially to Ivan Kostov for very valuable discussions.

\appendix

\section{Relative normalization of CFT and MQM couplings}
\label{A}

The relation between coupling constants can be obtained
from the requirement that the free energy of the perturbed MQM coincides with
the spherical partition function of the Sine--Liouville theory \Ref{conepert}.
The former has the following $\lambda$-expansion \cite{KKK}
\be
\CF_0(\mu,\lambda)=-{R\over 2}\mu^2 \log\mu -\lambda^2 \mu^{1/R}+O(\lambda^4).
\label{expF}
\ee
The first terms in the expansion of the Sine--Liouville partition function can be
calculated by integrating bulk three-point correlation functions
as follows
\beq
Z_{\rm sphere}&=&-\mathop{\lim}\limits_{b\to 1}\Biggl[
\int d\mul \int d\mul \int d\mul \langle V_{b}V_{b}V_{b}\rangle_{\rm sphere}
\Biggr. \nonumber \\
&+&\left. {\laml^2\over 2} \langl \cos^2(X/R)\rangl_{\rm sphere}
\int d\mul \langle V_{b-{1\over 2R}}V_{b-{1\over 2R}}
V_{b}\rangle_{\rm sphere}\] +O(\lambda^4).
\label{expZ}
\eeq
Three-point correlation functions in Liouville theory have been calculated in
\cite{DornXN,ZamolodchikovAA}. In particular, the three-point correlators
appearing in \Ref{expZ} are given by
\be
\langle V_{b}V_{b}V_{b}\rangle_{\rm sphere}=
b^{-1}\left[\pi\mul \right]^{1/b^2-2}[\gamma(b^2)]^{1/b^2}\gamma(2-1/b^2),
\label{threeb}
\ee
\be
\langle V_{b-{1\over 2R}}V_{b-{1\over 2R}}V_{b}\rangle_{\rm sphere}=
b^{-1}\left[\pi\mul \gamma(b^2)\right]^{{1\over b^2}+{1\over Rb}-2}
\gamma(b^2) \gamma\(b\(b-\oR\)\) \gamma\(2-{1\over b^2}-{1\over Rb}\).
\label{threel}
\ee
Integrating with respect to $\mul$ and neglecting non-universal $\mul^2$ terms,
one finds
\be
Z_{\rm sphere}=
-{1\over 2\pi^3}\,\muc^2 \log\muc -
{R\over 4\pi }\, \laml^2\muc^{1/R}\,
\gamma^2\(1-\oR\)
 +O(\lambda^4).
\label{expZint}
\ee
Comparing \Ref{expF} and \Ref{expZint}, one obtains
\be
{\mu \over \muc} =  {1 \over \sqrt{\pi^3 R}},
 \qquad
{\lambda \over \laml} = \sqrt{R\over4\pi}\(\pi^3 R\)^{1\over 4R}
\gamma\(1-\oR\) , \qquad
\({y \over \yl}\)^{{1\over 2R}-1}= {\pi R\over 2}\gamma\(1-\oR\),
 \label{relCFTMQM}
 \ee
where we introduced
$\yl=\muc\({1-R\over R^3}\,\laml^2\)^{-{R\over 2R-1}}$
similarly to the scaling parameter \Ref{scpt} in MQM.

\section{Two-point correlator from the matrix model}
\label{B}

In this appendix we generalize the results of section \ref{relMQMCFT} about the
matrix model curve $\SMQM$ to the second order in $\lambda$. From this generalization  
we also give some predictions concerning the complex curve $\SFZZ$ and two-point correlation functions.

The generalization of \Ref{ptau} to the next order reads as follows
\beq
x(\tau)&=&\sqrt{2\mu}\( \cos\tau +{\lambda\over R}\mu^{{1\over 2R}-1}
\cos\(\(1-\oR\)\tau\)
+{\(\oR -1\)\lambda^2 \over 2R^2}\mu^{\oR-2}\cos\tau\)+O(\lambda^3),
\label{xtau2}
\\
p(\tau)&=&-i\sqrt{2\mu}\( \sin\tau +{\lambda\over R}\mu^{{1\over 2R}-1}
\sin\(\(1-\oR\)\tau\)
+{\(\oR -1\)\lambda^2 \over 2R^2}\mu^{\oR-2}\sin\tau\)+O(\lambda^3).
\label{ptau2}
\eeq
The solution of the first equation with respect to $\tau$ is
\beq
& \tau(x)=\tau_0(x)+ {\lambda\over R}\, \mu^{{1\over 2R}-1}\,
{\cos\(\(1-\oR\)\tau_0(x)\)\over \sin \tau_0 (x)} &
\label{soltau2} \\
& +
{\(\oR -1\)\lambda^2 \over 2R^2}\mu^{\oR-2}\cot\tau_0(x)\(
1-2{\sin\(2\(\oR-1\)\tau_0(x)\) \over \sin\(2\tau_0(x)\)}-
{R\over 1-R} {\cos^2\(\(\oR-1\)\tau_0(x)\) \over \sin^2\tau_0(x)}\)+
O(\lambda^3). &
\nonumber
\eeq
Substituting this solution into \Ref{ptau2}, one obtains
\beq
& p(\tau_0)=-i\sqrt{2\mu}\[\sin\tau_0+{\lambda\over R}\, \mu^{{1\over 2R}-1}\,
{\cos{\tau_0\over R}\over \sin \tau_0 }
\right. & \label{pofx2} \\
&- \left.{\lambda^2 \mu^{\oR-2}\over 2R^2\sin\tau_0}\(
{\cos^2\(\(\oR-1\)\tau_0\) \over \sin^2\tau_0}+
\(\oR-1\){\sin\(\({2 \over R}-1\)\tau_0\) \over \sin\tau_0}\)\]
+O(\lambda^3). &
\nonumber
\eeq
This equation describes the complex curve $\SCFT_1$ up to the second order in the
Sine--Liouville perturbation. This order corresponds already to the two-point correlation
function of the Sine--Liouville operator
\be
\CT=\int d^2\sigma\, \cos(X/R)\,V_{1-{1\over 2R}}.
\label{SLop}
\ee

Now we show how this correlator can be extracted from the curve \Ref{pofx2}.
First, using the property \Ref{ZZnZZ} and the identifications \Ref{CCC} and \Ref{identpar},
one can find all functions $\rho_{(n)}$
\beq
&\rho_{(n)}(\ss)=-\CC^{-1}\sqrt{2\mu}
\[ n\, \sinh\(\pi\ss\) -{\lambda\over R}\, \mu^{{1\over 2R}-1}\,
 {\sin\({\pi n\over R}\)\over \sin {\pi \over R}}
 {\cosh{\pi\ss\over R}\over \sinh \(\pi\ss\)}
- {\lambda^2 \mu^{\oR-2}\over 4R^2\sinh^3\(\pi\ss\)} \biggl\{
n  +
 \biggr. \right. & \label{rhonnn} \\
& \left.  \left. {\sin\({2\pi n\over R}\)\over \sin {2\pi \over R}}
\( \cosh\(2\(\oR-1\)\pi\ss\) -2\(\oR-1\)\sinh\(\pi\ss\) \sinh\(\({2 \over R}-1\)\pi\ss\)\)
\right\}\]
+O(\lambda^3), &
\nonumber
\eeq
where we kept MQM couplings to avoid huge coefficients.
This result allows to restore the full function $w(\ss)$, describing the curve $\SFZZ$,
up to a periodic term. It is given by
\beq
& w(\ss;\lambda)=
-\pi \CC^{-1}\sqrt{2\mu}
\[ \ss\, \sinh\(\pi\ss\) -{\lambda\over R}\, \mu^{{1\over 2R}-1}\,
 {\sinh{\pi\ss\over R}\over \sin {\pi \over R}\sinh\(\pi\ss\)  }
\right.  & \label{wfull} \\
&- \left. {\lambda^2 \mu^{\oR-2}\over 4R^2} \left\{{1\over\sinh^3 \(\pi\ss\) }
\(\ss + {\sinh\(2\(\oR-1\)\pi\ss\)\over \sin {2\pi \over R}}  -
2\(\oR-1\){\sinh \(\pi\ss\)\over \sin {2\pi \over R}}\cosh\(\({2 \over R}-1\)\pi\ss\)\)
-g(\ss)\right\}
\]
 +O(\lambda^3).
& \nonumber
\eeq
where $g(\ss+2i)=g(\ss)$. We also require that $w$ be a symmetric function of $\ss$
as it is in the non-perturbed case.
Hence, we have the additional constraint $g(-\ss)=g(\ss)$. This fixes $g$ to be
a meromorphic function of $\mubc$: $g(\ss)=\tilde g(\cosh(\pi\ss))$. 
Although such contribution seems to be trivial,
it can obtain non-trivial analyticity properties after integration
with respect to $\mubc$. Performing the integration, one obtains
\beq
& \ZFZZ(\ss;\lambda)=
-{2C^{-1}}\mu
\[ {1\over 4\pi} \( \pi\ss \sinh\(2\pi\ss\)-{1\over 2} \cosh\(2\pi\ss\)-
\(\pi\ss\)^2\) -
{\lambda}\, \mu^{{1\over 2R}-1}\,
 {\cosh{\pi\ss\over R}\over \sin {\pi \over R}  }
\right.  & \label{Zfull} \\
&+ \left. {\lambda^2 \mu^{\oR-2}\over 4R^2} \(
\ss \coth\(\pi\ss\)-{1\over \pi}\log\sinh\(\pi\ss\)
+ {\sinh\(\({2 \over R}-1\)\pi\ss\)\over \sin {2\pi \over R}\sinh\(\pi\ss\)} 
+\int \tilde g(\cosh(\pi\ss))\,d\cosh(\pi\ss)
\) \]
 +O(\lambda^3).
& \nonumber
\eeq
The last term in \Ref{Zfull} gives (one half of) the two-point correlation function
of the Sine--Liouville operator $\CT$ \Ref{SLop} on the disk with the Neumann boundary conditions
on the Liouville field and the Dirichlet conditions on the matter field $X$.
Rewriting it in terms of the CFT couplings by means of \Ref{relCFTMQM}, one finds
the following prediction of MQM
\beq
&\langl \CT^2 \rangl_{\ss,{\rm Dir}}=-{\pi \DD\over 2^{5/4}\sqrt{R}}\,\laml^2 \muc^{\oR-1}
\gamma^2\(1-\oR\)
\( \ss \coth\(\pi\ss\)
+ {\sinh\(\({2 \over R}-1\)\pi\ss\)\over \sin {2\pi \over R}\sinh\(\pi\ss\)}
\right. & \label{predtwo} \\
& \biggl. +\int \tilde g(\cosh(\pi\ss))\,d\cosh(\pi\ss) 
-{1\over \pi}\log\sinh\(\pi\ss\) \biggr). &  
\nonumber
\eeq
We grouped the logarithmic term together with the integral of $\tilde g$ because, in fact, it
can be canceled by an appropriate choice of this function.
For example, it is enough to take $\tilde g(x)={1\over \pi}{x\over x^2-1}$.
Unfortunately, the function $\tilde g$ is not fixed by our approach.
But since the logarithmic singularity is quite unexpected, 
most likely, it is indeed canceled by $\tilde g$.
On the other hand, the poles of the first two terms in \Ref{predtwo} are unavoidable.
Thus, in contrast to the one-point bulk correlator, the two-point correlation function
has pole singularities at the values of the parameter $\ss$ corresponding to the $(m,n)$ branes.

\section{Singularities of $\SCFT_n$}
\label{C}

For all quantities we encountered, the perturbative expansion in the Sine--Liouville 
coupling breaks down near the singularities of the Riemann surfaces when the quantities
are considered as functions of the parameter $\tau_0$.
The reason for that can be traced out to the divergence of the multi-point correlation functions
with the FZZ boundary conditions at the points $\tau_0=\pi k$. 
As a result, the perturbative expansion does not allow to find the positions of
the $(m,n)$-branes associated to the singularities. 
In other words, we cannot solve the equation $\rho_{(n)}=0$.

A way to avoid this obstacle would be to find a parameter in terms of which
all relevant functions are analytic near the singularities and have a well defined
expansion in $\lambda$.
In the case of the Riemann surface $\SCFT_1$, such a parameter is given  
by the Euclidean time $\tau$. Therefore, it is natural first to try to use this parameter also
for $n>1$. 
To express $\rho_{(n)}$ through $\tau$, one should 
invert the relation \Ref{soltau2} and then substitute the result into \Ref{rhonnn}.
The first step gives
\be
\tau_0=\tau- {\lambda\mu^{{1\over 2R}-1}\over R}\,
{\cos\(\(1-\oR\)\tau\)\over \sin \tau} -
{\lambda^2 \mu^{\oR-2}\over 2R^2}\cot\tau\(\oR -1
+{\cos^2\(\(\oR-1\)\tau\) \over \sin^2\tau}\)+
O(\lambda^3). 
\label{soltauzero}
\ee
The substitution of this expansion  into \Ref{rhonnn} leads to the
following horrible expression
\beq
\rho_{(n)}(\tau)&=&-i\CC^{-1}\sqrt{2\mu}
\Bigl[ n\, \sin\tau
\Bigr.\label{rhotaumn} \\
&-&{\lambda\over R}\, \mu^{{1\over 2R}-1}
\( \sin\(\(\oR-1\)\tau\)+\(1- {\sin\({\pi n\over R}\)\over n \sin {\pi \over R}} \)
 {\cos{\tau\over R}\over \sin\tau}\)
\nonumber \\
& - &  {n\lambda^2 \mu^{\oR-2}\over 2R^2\sin^3\tau} \(
\(1-{\sin\({\pi n\over R}\)\over n\sin {\pi \over R}}\) +
\(\oR-1\)\(\cos^2\tau-{\sin\({\pi n\over R}\)\over n\sin {\pi \over R}}\)
\right. \nonumber \\
&& \qquad
+\hf \(1-\(\oR+1\) {\sin\({\pi n\over R}\)\over n\sin {\pi \over R}}+\oR
 {\sin\({2\pi n\over R}\)\over n\sin {2\pi \over R}} \) \cos\(2\(\oR-1\)\tau_0\)
\nonumber \\
&&\left.\left.  \qquad
+\hf \(\oR-1\){\sin\({\pi n\over R}\)\over n\sin {\pi \over R}}
\(1-{\cos\({\pi n\over R}\)\over \cos {\pi \over R}}\)
\cos{2\tau\over R}
\) \]
+O(\lambda^3).
\nonumber
\eeq
The main feature of this result is that each order in $\lambda$ contains terms
proportional to $\lambda^k\(\sin\tau\)^{1-2k}$ similarly to \Ref{rhonnn}.
This shows that for $n>1$ the expansion of $\rho_{(n)}(\tau)$ does not differ qualitatively 
from the expansion of $\rho_{(n)}(\tau_0)$ and, hence, 
the parameter $\tau$ does not provide a well defined expansion.

Thus, one should look for another parameter $\ttau(\tau_0)$. 
The minimal requirement would be that 
the largest singularity in the $k$th order in
the expansions of $x(\ttau)$ and $\rho_{(n)}(\ttau)$ is $\(\sin\ttau\)^{-k}$.
Then both expansions are well defined provided that 
near a singularity of $\Sigma_n$ the parameter $\ttau$ behaves as 
$\ttau_{m,n}=\pi m+O(\sqrt{\lambda})$. 
However, we see that the number of conditions is twice more than the number of 
free coefficients in $\ttau(\tau_0)$.
Therefore, we do not expect that there exists a solution of this problem.
If this is indeed the case, 
this means that there is no well defined coordinate parameterizing the Riemann surface $\SCFT_n$
near the zeros of $\rho_{(n)}$ and we remain with two possibilities: either the zeros correspond 
to essential singularities or there are no zeros, and consequently there are no singularities at all.
Thus, our conclusion is that most likely after the perturbation by the Sine--Liouville 
operator the complex curves $\SCFT_n$ with $n>1$ 
do not have singularities (except the trivial one with $m=0$) and the $(m,n)$ ZZ branes
do not appear in the perturbed $c=1$ string theory.

\end{document}